\documentclass[%
 reprint,
 amsmath,amssymb,
 aps,
 prc,
floatfix
]{revtex4-2}

\usepackage{graphicx}% Include figure files
\usepackage{dcolumn}% Align table columns on decimal point
\usepackage{bm}% bold math

\usepackage{array}
\usepackage{siunitx}
\usepackage{adjustbox}
\usepackage{comment} 

\graphicspath{{./figures/}} %Where the figures folder is located

\begin{document}

\preprint{APS/123-QED}

\title{Constraining the destruction rate of $^{40}$K in stellar nucleosynthesis through the study of the $^{40}$Ar(p,n)$^{40}$K reaction}

% Force line breaks with \\

\author{P.~Gastis}
     \email{gasti1p@cmich.edu}

\author{G.~Perdikakis}%
     \email{perdi1g@cmich.edu}

\author{J.~Dissanayake}%
\author{P.~Tsintari}%
\author{I.~Sultana}

\affiliation{%
Department of Physics, Central Michigan University, Mt. Pleasant, MI 48859, USA\\
Joint Institute for Nuclear Astrophysics - Center for the Evolution of the Elements, East Lansing, MI 48824, USA
}%

\author{C.R.~Brune}%
%\author{D. Carter}%
\author{T.N.~Massey}%
\author{Z.~Meisel}%
\author{A.V.~Voinov}%

\author{K. Brandenburg}
\author{T.~Danley}
\author{R.~Giri}
\author{Y.~Jones-Alberty}
\author{S.~Paneru}
\author{D.~Soltesz}
\author{S.~Subedi}

\affiliation{%
Department of Physics \& Astronomy , Ohio University, Athens, OH 45701, USA\\
Joint Institute for Nuclear Astrophysics - Center for the Evolution of the Elements, East Lansing, MI 48824, USA
}%



\date{\today}% It is always \today, today,

%---------------------
%    ABSTRACT
%---------------------

\begin{abstract}

\textbf{Background:} $^{40}$K plays a significant role in the radiogenic heating of earth-like exoplanets, which can affect the development of a habitable environment on their surfaces. The initial amount of $^{40}$K in the interior of these planets depends on the composition of the interstellar clouds from which they formed. Within this context, nuclear reactions that regulate the production of $^{40}$K during stellar evolution can play a critical role. \textbf{Purpose:} In this study, we constrain for the first time the astrophysical reaction rate of $^{40}$K(n,p)$^{40}$Ar, which is responsible for the destruction of $^{40}$K during stellar nucleosynthesis. We provide to the nuclear physics community high-resolution data on the cross-section and angular distribution of the $^{40}$Ar(p,n)$^{40}$K reaction. These are important to various applications involving $^{40}$Ar. The associated reaction rate of the $^{40}$Ar(p,n)$^{40}$K process addresses a reaction rate gap in the JINA REACLIB database in the region of intermediate-mass isotopes \textbf{Methods:} We performed differential cross-section measurements on the $^{40}$Ar(p,n)$^{40}$K reaction, for six energies in the center-of-mass between 3.2 and 4.0~MeV and various angles between 0$^{\circ}$ and 135$^{\circ}$. The experiment took place at the Edwards Accelerator Laboratory at Ohio University using the beam swinger target location and a standard neutron time-of-flight technique. We extracted total and partial cross-sections by integrating the double differential cross-sections we measured. 
\textbf{Results:}
 The total and partial cross-sections varied with energy due to the contribution from isobaric analog states and Ericson type fluctuations. The energy-averaged neutron angular distributions were symmetrical relative to 90~$^\circ$. Based on the experimental data, local transmission coefficients were extracted and were used to calculate the astrophysical reaction rates of $^{40}$Ar(p,n)$^{40}$K and $^{40}$K(n,p)$^{40}$Ar reactions. The new rates were found to vary significantly from the theoretical rates in the REACLIB library. We implemented the new rates in network calculations to study nucleosynthesis via the slow neutron capture process, and we found that the produced abundance of $^{40}$K is reduced by up to 10\% compared to calculations with the library rates. At the same time, the above result removes a significant portion of the previous theoretical uncertainty on the $^{40}$K yields from stellar evolution calculations. \textbf{Conclusions:} Our results support that the destruction rate of $^{40}$K in massive stars via the $^{40}$K(n,p)$^{40}$Ar reaction is larger compared to previous estimates. The rate of $^{40}$K destruction via the $^{40}$K(n,p)$^{40}$Ar reaction now has a dramatically reduced uncertainty based on our measurement. This result directly affects the predicted stellar yields of $^{40}$K from nucleosynthesis, which is a critical input parameter for the galactic chemical evolution models that are currently employed for the study of significant properties of exoplanets.

%\begin{description}
%\item[Usage] Secondary publications and information retrieval purposes.
%\item[Structure] You may use the \texttt{description} environment to structure your abstract; use the optional argument of the \verb+\item+ command to give the category of each item. 
%\end{description}

\end{abstract}

\maketitle
%\tableofcontents

%---------------------
%    INTRODUCTION
%---------------------

\section{\label{sec_intro}INTRODUCTION}

$^{40}$K is a very long-lived naturally occurring radioisotope of potassium. The slow beta decay rate that transforms it into $^{40}$Ar gives it important roles beyond that of the enrichment of the interstellar medium. One such role is in the area of nucleocosmochronology~\cite{Cowan1991}. A possibly more exciting role, however, is related to the evolution of habitable environments in earth-like extrasolar planets. $^{40}$K is typically found in the interiors of exoplanets, a remnant of the nucleosynthesis events that enriched the interstellar gas cloud and, in turn, gave birth to the main-sequence star that such exoplanets typically orbit. The decay of $^{40}$K is an exothermic process that generates heat. It is hence found among other radioactive elements responsible for keeping an Earth-like planet's mantle hot for the billions of years following its birth. This so-called radiogenic heating and its evolution since a planet's formation are critically connected to the initiation and sustainability of any tectonic activity of the earth-like planet as well as of other planetary functions that control CO$_2$ levels in a planet's atmosphere. Processes that affect the emission and absorption of greenhouse gases on a planet control the balance of heat and influence habitability \cite{Foley2018a}. It has been shown that it is the initial composition of a planet in long-lived radioactive elements and particularly the quantities upon the formation of $^{40}$K and $^{235}$U that are critical parameters in the evolution towards a habitable environment \cite{Foley2018a, Frank2014a}.

\begin{figure}
\includegraphics[width=0.40\textwidth]{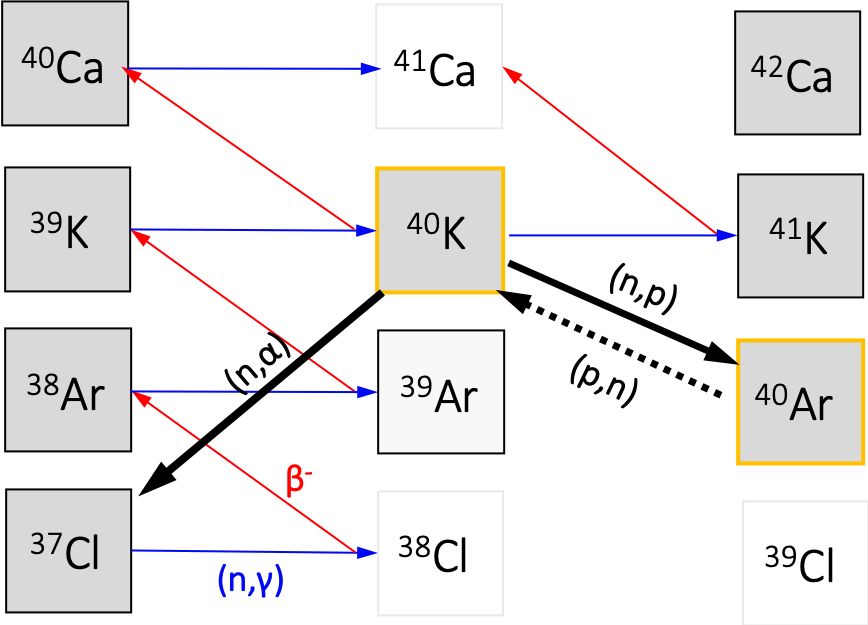}
\caption{Flow diagram of nuclear reactions around $^{40}$K during s-process nucleosynthesis. The main avenues for destruction of $^{40}$K are through the $(n,p)$ and $(n,\alpha)$ reaction rates (solid black arrows). In this work we infer the $^{40}$K(n,p)$^{40}$Ar rate from a study of the $^{40}$Ar(p,n)$^{40}$K reaction (dashed black arrow).}
\label{fig_flow}
\end{figure}

In massive stars, $^{40}$Ar and $^{40}$K are mainly created through neutron capture reactions. The abundances of the two isotopes are connected since they share a common reaction flow during helium, carbon, and neon burning. Nucleosynthesis calculations have shown that the abundances of the two isotopes are mutually sensitive to the reaction rates responsible for their destruction~\cite{Hoffman1999}. For $^{40}$K these rates are the $^{40}$K(n,$\alpha$)$^{37}$Cl and $^{40}$K(n,p)$^{40}$Ar reactions. An example of a reaction flow involving these rates in the context of a simple model of weak s-process nucleosynthesis is shown in Fig.~\ref{fig_flow}. $^{40}$Ar is converted back to $^{40}$K via the $^{40}$Ar(p,n)$^{40}$K reaction, a process that needs, however, temperatures of the order of 5~GK to switch on.

A search in the JINA REACLIB library \cite{Cyburt_2010} --we used REACLIB v2.2-- commonly used for astrophysics applications reveals a shocking lack of experimental data on (p,n) and (n,p) reactions for intermediate-mass nuclei with atomic numbers between Z=14 and Z=33 and 30$<$A$<$70. The only available experimental data in the library are those for $^{41}$K(p,n)$^{41}$Ca and its time-inverse rate for $^{41}$Ca(n,p)$^{41}$K. A detailed study of the $^{40}$Ar(p,n)$^{40}$K reaction at energies into -or close to- the Gamow window allows to set constraints on both the direct and the inverse astrophysical reaction rates and to improve the accuracy of relevant abundance predictions for the astrophysics applications mentioned above.

$^{40}$Ar is an increasingly more valuable gas used in various scientific fields. In the last two decades, $^{40}$Ar has found extended use in neutrino physics experiments due to its low cost, non-reactive nature, and relatively easy purification process. Purified $^{40}$Ar, is used in Liquid Argon Time Projection Chambers (LArTPCs) which are integrated in neutrino tracking detector systems such as MicroBooNE~\cite{Acciarri:2016smi}, ICARUS~\cite{Amerio:2004ze}, and LArIAT~\cite{Cavanna:2014iqa}. These systems are optimized for detecting accelerator, solar, and supernovae neutrinos for experiments relevant to neutrino physics and astrophysics. $^{40}$Ar has used in accelerator physics technological applications. Such uses of $^{40}$Ar include its placement in stripping gas-targets in tandem accelerators~\cite{Makarov2015}, and in testing stable beams in low- and high-energy linear accelerator facilities such as ReA3 at the Facility of Rare Isotope Beams (FRIB) and Linac3 at CERN. 

In the context of the last two applications, studies on proton and neutron-induced reactions on $^{40}$Ar at MeV-scale energies, are useful in the development of diagnostics tools, for testing experimental setups~\cite{Gastis:2017bht}, and for background characterization~\cite{Bhike2014} in long-baseline experiments. Currently, the only available experimental data for the $^{40}$Ar(p,n)$^{40}$K reaction at energies below 20~MeV, come from low energy neutron yield measurements for the determination of reaction thresholds, Q-values, and Isobaric Analog States~\cite{richards1948,parks1966,holland1959,Young1968a}.  The energy resolution of some of these earlier works is comparable to the data presented here, but the absolute quantification of the cross-section was not a primary goal of the authors. The work of Young et al.~\cite{Young1968a} deserves particular mention, since it included an attempt at the measurement of the total cross-section using a long counter. While a plot of the excitation function was given, the authors acknowledged the non-quantified systematic uncertainty in the energy dependence of their neutron detector's efficiency and did not tabulate the cross-section. Presumably, the unknown energy dependence of the efficiency is the reason that the work of Young et al. is not included in any database of cross-section data for the (p,n) reaction.   

In this work, we present for the first time a complete set of high-resolution experimental data on the $^{40}$Ar(p,n)$^{40}$K reaction excitation function and its angular distribution. We extract the absolute total cross-section, as well as partial cross-sections of the $^{40}$Ar(p,n$_x$)$^{40}$K, (x=0,1,2) reactions to the ground and two first levels of excitation of $^{40}$K. We compare the cross-sections and the angular distributions with the theory and find that the results are consistent with the predictions of the statistical model. In particular, we explain the details of the excitation function in the context of the phenomenon of isobaric analog states and Ericson-type fluctuation theory, which are applicable in our case. We use the experimental data to extract local values for the proton - $^{40}$Ar transmission coefficients. From the experimental cross-section data, we calculate the experimentally constrained astrophysical reaction rate for the $^{40}$Ar(p,n)$^{40}$K reaction. Using detailed balance, we extract the reaction rate for the destruction of $^{40}$K via the $^{40}$K(n,p)$^{40}$Ar and find it significantly different from the recommended library rate that is based on theory. Finally, by implementing the extracted rates in network calculations, we study s-process nucleosynthesis and find that the produced abundance of $^{40}$K is $\sim$10\% lower compared to calculations with the REACLIB rates. 

We have divided the article into four sections. In Section~\ref{sec_setup}, we describe the experimental setup that we used for the measurement of differential cross sections and angular distributions at Ohio University. In Section~\ref{sec_analysis}, we discuss the details of the data analysis and the extraction of neutron efficiencies, while in Section~\ref{sec_results}, we summarize the results and compare them with the statistical model calculations. In addition, we discuss the implications of this study in s-process nucleosynthesis and show results from network calculations. The conclusions of our work are summarized in Section~\ref{sec_conclusion}.

%% ----------------------------------------
%%           EXPERIMENTAL SETUP
%% ----------------------------------------
\section{\label{sec_setup}Experimental Setup}

\begin{figure}
\includegraphics[width=0.55\textwidth]{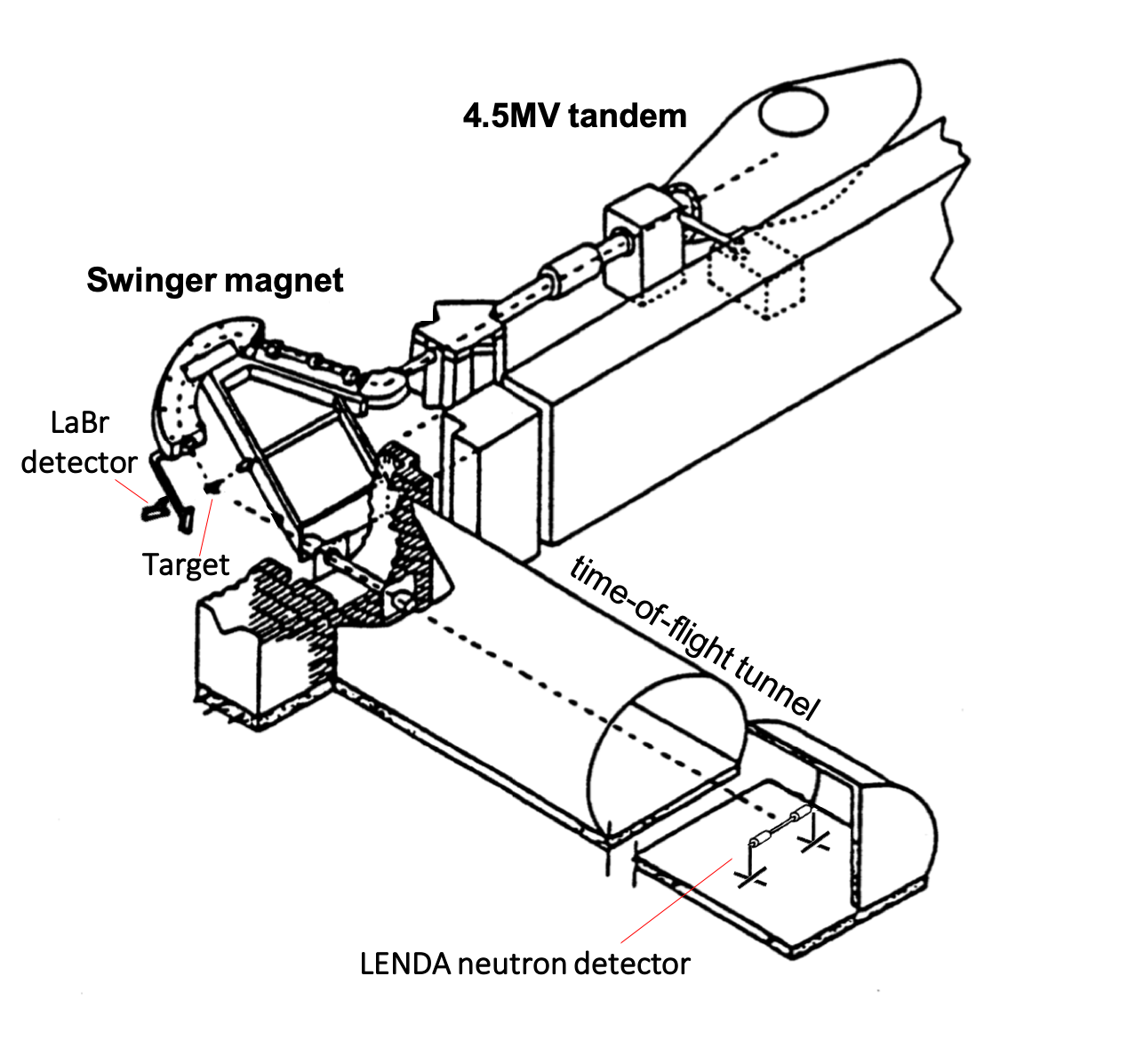}
\caption{Diagram of the beam swinger and of the neutron time-of-flight tunnel at the Edwards Accelerator Laboratory (modified image from \cite{meisel_caari}). The beam swinger allows the beam axis to rotate around the target center, making possible measurements of neutron yields at various reaction angles without the need for multiple neutron flight paths. The swinger is part of the 4.5~MV Tandem accelerator of Ohio University.}
\label{fig_setup}
\end{figure}

The experiment was carried out at the 4.5~MV tandem accelerator of the Edwards Accelerator Laboratory using the beam swinger capability of the facility  \cite{meisel_caari}. Pulsed proton beams at energies between 3.5 and 4.2~MeV were delivered from the tandem accelerator into a cylindrical gas-cell target at the end of the swinger magnet, as shown in Fig~\ref{fig_setup}. %0.25mm-thick stainless steel walls
The cylindrical body of the gas-cell was 3~cm in length and had a 9~mm internal diameter. A set of beam collimators upstream of the gas-cell provided a beam spot of $\sim$4~mm in diameter at the target position, while a 1~mm-thick gold layer at the bottom of the cylinder served as a beam stopper. To keep the beam-induced neutron background as low as possible, the entrance window of the gas-cell was made out of high purity ($\sim$99.5\%) aluminum and was 10.37$\pm$0.04 $\mu$m-thick. The thickness and composition of the aluminum foil were determined via Rutherford back-scattering spectroscopy (RBS) and proton-induced x-ray emission (PIXE) analysis at Ohio University. During the data taking phase of the experiment, the gas-cell was filled with natural argon gas (99.999\%) and was pressurized at 198$\pm$5~Torr.  This pressure corresponded to an areal density of (1.97$\pm$0.01)$\cdot 10^{19}$~atoms/cm$^2$ for $^{40}$Ar, while the purity of the gas was verified with a residual gas analyzer (RGA) at the end of the experiment. Lower (98~Torr) and higher (414~Torr) pressures than the one for data taking were also used to understand the qualitative features of the cross-section and its fluctuating character. The use of various pressures and, consequently, target thicknesses allowed us to interpret the neutron spectra in terms of the interplay between energy resolution of the experiment and fluctuations in the cross-section.

For the detection of neutrons, we used a plastic scintillator detector that is similar to the Low Energy Neutron Detector Array (LENDA) bars~\cite{Perdikakis:2011yf}. The size of the detector's crystal was 31$\times$4.5$\times$2.5~cm$^3$ (Length $\times$ Height $\times$ Width), which is 1~cm longer compared to a standard LENDA bar. We placed the bar inside the neutron time-of-flight tunnel at a distance of 5~m from the gas target. The swinger magnet, which can rotate around the center of the target at angles between 0~$^\circ$ $\leq \theta_{\rm lab} \lesssim$ 180~$^\circ$, defined the angle $\theta_{\rm lab}$ of the detector relative to the beam axis. During the experiment, the neutron detector was oriented horizontally relative to the ground (see Fig~\ref{fig_setup}). Having the crystal in this position, we were able to achieve an angular resolution of 0.25~$^\circ$ in $\theta_{\rm lab}$. The uncertainty in timing due to the crystal's length was less than 0.5~ns. The intrinsic timing resolution of LENDA is also less than 0.5~ns \cite{Perdikakis:2011yf}. Therefore, the pulsing and bunching of the beam dominated the timing uncertainty, which we measured using the neutron spectra to be no more than 2~ns.

For determining the neutron angular distributions, we performed measurements at various angles between 0~$^{\circ}$ and 135~$^{\circ}$. While the swinger magnet can rotate up to 180~$^{\circ}$, the system's supporting materials upstream of the gas-cell significantly shadowed the target for $\theta_{lab}$~$>$~135~$^{\circ}$. For this reason, we did not use in this study any measurements at larger angles than 135~$^{\circ}$.

To determine the intrinsic efficiency of LENDA at the neutron energies of the experiment, we performed an efficiency calibration measurement using the neutron spectrum of the $^{9}$Be(d,n) reaction for a thick Be-target. The $^{9}$Be(d,n) reaction is a suitable and well-characterized source of neutrons for calibrating neutron detectors in the energy range 0.1 $\leq E_{n}\leq$ 11.7~MeV due to its high neutron flux at the low- and high- energy ends \cite{howard_be,massey_al,Meadows:1991}. In this measurement, we used a 7~MeV deuterium beam at $\theta_{\text{lab}}$=0~$^{\circ}$, while the position of LENDA remained the same as described above.

In addition to the neutron detector, a 4.5$\times$5.0~cm LaBr$_{3}$:Ce scintillator was used to detect gamma rays associated with the $^{40}$Ar(p,n) reaction. The detector was mounted at 90~$^\circ$ relative to the beam axis and was 18~cm away from the target. Gamma spectra from the LaBr detector and particularly the yields of the gamma transitions associated with the excited levels of $^{40}$K were used for qualitative analysis and cross-verification of the neutron detector results. These data were invaluable in comprehending the detailed characteristics of the neutron spectra, as described in the analysis section. 

For getting a detailed picture of the photon emission as a function of proton energy, the gamma-ray data were collected separately from the time-of-flight measurements. In this way, we were able to increase the energy resolution by lowering the gas pressure in the cell and take smaller energy steps. Twenty-five spectra were collected in total for beam energies between 3.4 to 4.0~MeV (with 25~keV step), while the pressure in the gas-cell was 98$\pm$5~Torr.

%% ----------------------------------------
%%            ANALYSIS
%% ----------------------------------------

\section{\label{sec_analysis}Data Analysis}

We extracted the total cross-section of the $^{40}$Ar(p,n)$^{40}$K by summing all the partial cross-sections of the $(p,n_x)$ exit channels populating the ground and excited levels of $^{40}$K. To obtain the partial cross-sections, we followed the procedures outlined below to integrate the experimental double differential cross sections over the angle $\theta$. 

\begin{figure}[h]
\includegraphics[width=0.45\textwidth]{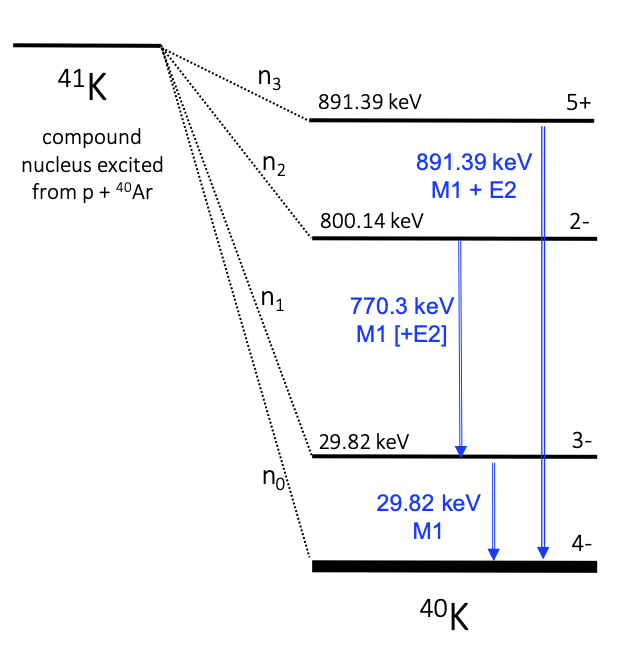}
\caption{Level scheme of excited levels in the final $^{41}$K residual of the $^{40}$Ar(p,n)$^{40}$K reaction (Q = -2.2868~MeV). The $^{41}$K compound nucleus excitation energy ranged between 11.2 - 11.7~MeV. The scheme shows the energetically allowed excited states in the residual nucleus along with the prominent gamma-rays from the de-excitation of the residual nucleus. The corresponding neutron energies span between 1.0--1.6~MeV for n$_0$, n$_1$, and 0.1--0.9~MeV for n$_2$, n$_3$.}
\label{fig_level}
\end{figure}

Fig.~\ref{fig_level} shows schematically the portion of the $^{40}$K level scheme that is relevant for our study at energies between 3.0 and 4.0~MeV in the center-of-mass system along with the associated neutron transitions n$_{0-3}$ assuming a reaction proceeding via the formation of a $^{41}$K compound nucleus. The kinetic energies of the emitted neutrons from the various $(p,n_x)$ channels are expected to range between 0.1 and 1.6~MeV.

Within the limits of the timing resolution of the system as well as the energy resolution of the incident beam, the peaks from the $(p,n_{0})$ and $(p,n_{1})$ channels were not expected to be resolved in the neutron spectra due to their small difference in neutron time-of-flight. As a result, cumulative angular distributions and partial cross-sections were obtained for the $(p,n_{0,1})$ channels. Neutron peaks in our spectra corresponding to these two channels spanned the energy range of  1.0$\lesssim E_{n_{0},n_{1}} \lesssim$1.6~MeV.

For the $(p,n_{2})$ and $(p,n_{3})$ channels, the resolution of the system was adequate to separate their neutrons that were expected in the combined energy range of 0.1$\lesssim E_{n_{2},n_{3}} \lesssim$0.9~MeV.  The fact that we observed no neutrons (or gamma rays) from the $(p,n_{3})$ channel suggests that the corresponding partial cross-section was too low compared to the sensitivity of our experiment.
Within our experimental uncertainties, the contribution of the $(p,n_3)$ channel to the total cross-section was considered negligible compared to the dominant $(p,n_2)$ channel.

%% ----------------------------------------
%%           ANGULAR DISTRIBUTIONS
%% ----------------------------------------

The differential cross sections at angle $\theta$ and energy E were calculated using the following relation: 
\begin{equation}\label{eq_dsigma}
    \frac{d\sigma}{d\Omega}(\theta,E) = \frac{ I_n }{\tau_{d} \epsilon_{n} N_{t}N_{p}\Delta\Omega}
\end{equation}
where $I_n$ is the number of detected neutrons, $\tau_{d}$ and  $\epsilon_{n}$ are correction factors for the dead-time of the electronics system and the intrinsic efficiency of LENDA, respectively, $N_t$ is the areal density of the target nuclei, $N_p$ is the total number of beam particles impinging the target during the measurement, and $\Delta\Omega$ is the solid angle of the detector. The number of detected neutrons $I_n$ was extracted from the time-of-flight spectra after applying particular thresholds levels in the output pulse heights of the detector. By doing this, we were able to obtain the corresponding neutron efficiencies $\epsilon_{n}$ from the LENDA efficiency data (see Section~\ref{sec_neutr_eff}). 

By collecting at least four data points for each angular distribution and converting the data to the center-of-mass system, a least-squares fit was then applied using Legendre polynomial expansions of the form:
\begin{equation} \label{legendre}
\left.\frac{d\sigma}{d\Omega}\right\rvert_{CM} = \sum_{i=0}^{n} a_{i}P_{i}(cos\theta_{cm}) 
\end{equation}
where P$_{i}$ is the i$^{th}$ order Legendre polynomial. Having properly reproduced the angular distributions in the range from 0~$^{\circ}$ to 180~$^{\circ}$, the total cross sections were obtained by integrating the fit functions. 

\begin{figure*}
\includegraphics[width=1.0\textwidth]{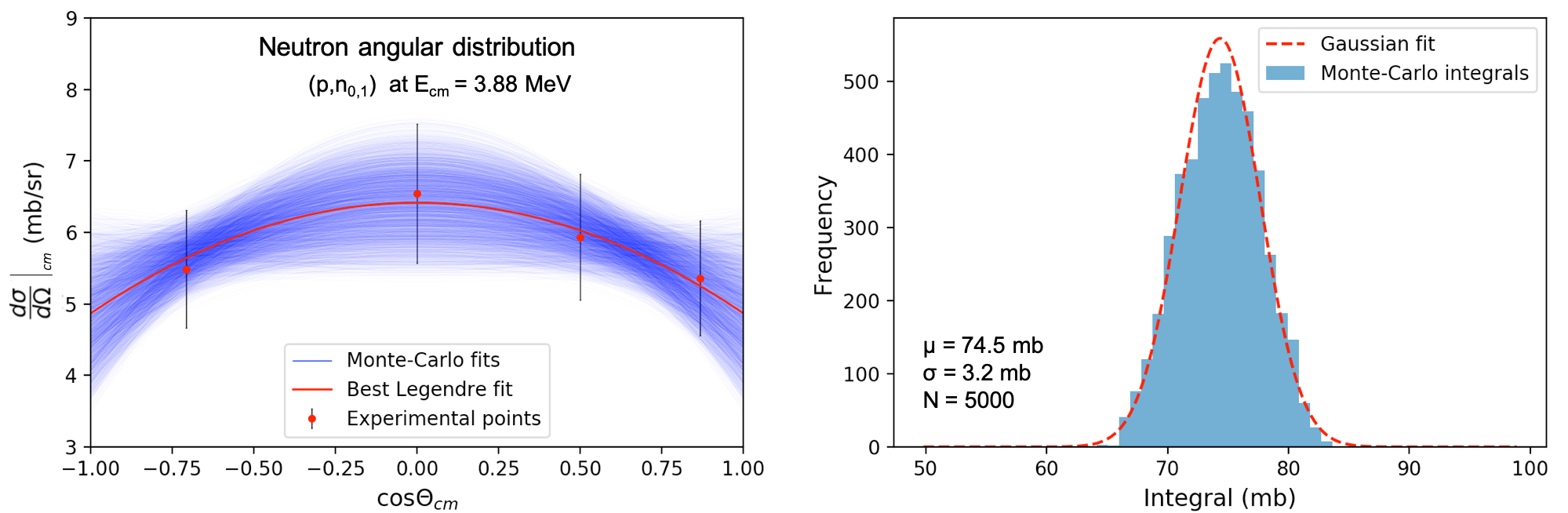}
\caption{Calculation of the total cross-section from neutron angular distributions and corresponding uncertainties. The Monte-Carlo method described in the text was used to estimate these uncertainties from the spread of results for the integrated angular distribution fits. 
(\textbf{Left}) Example of a measured angular distribution of the $(p,n_{0,1})$ channel at E$_{CM}$=3.88~MeV. The red line shows the best Legendre fit on the experimental points, while each one of the 5000 blue lines, corresponds to a fit of a randomly chosen configuration of points within the limits of each point's statistical uncertainty. (\textbf{Right}) Distribution of the angle-integrated cross-sections that were obtained from N sampled fits. The mean value $\mu$ and standard deviation $\sigma$ of the distribution, are extracted from the Gaussian fit of the full distribution. The final cross-sections are given as $\mu \pm 3\sigma$ (Table \ref{tab:results_sigma}).}
\label{fig_ang_errors}
\end{figure*}

The total cross-section values and associated uncertainties were estimated using a Monte-Carlo sampling technique that took into account the experimental errors of the differential cross-section measurements. In this analysis, the angular distribution data points were randomly varied within the limits of their error-bars. For each randomly chosen configuration of points, an integrated angular distribution was obtained. By repeating this process N-times, we were able to create a distribution of angle-integrated cross-sections where its standard deviation ($\sigma$) and mean value ($\mu$) were used to determine the final values of the cross-section and its error. In order to assure that all the possible fit functions are taken into account in the estimate of the final errors, the data points were varied according to a flat random distribution.
Fig.~\ref{fig_ang_errors} shows an example of an angular distribution that was analyzed with the Monte-Carlo method we describe here. All the distributions were fitted using Legendre polynomials up to the second order.
In cases of symmetrical distributions, only the even terms of the expansion were included (e.g., Fig.~\ref{fig_ang_errors}). Any uncertainties associated with the integration method on the fit functions were negligible.

In the following subsections, we provide details on the analysis of the neutron efficiency necessary for the quantification of the cross-section in Eq. \ref{eq_dsigma}. We also describe the analysis of gamma-ray spectra used to extract relative gamma-ray yields and to interpret the cross-section results qualitatively.

%% ----------------------------------------
%%           NEUTRON EFFICIENCY
%% ----------------------------------------
\subsection{\label{sec_neutr_eff}Neutron Efficiency} 

\subsubsection{\label{subsec_eff_curve}Extraction of the detector efficiency curve} 

\begin{figure}
\includegraphics[width=0.51\textwidth]{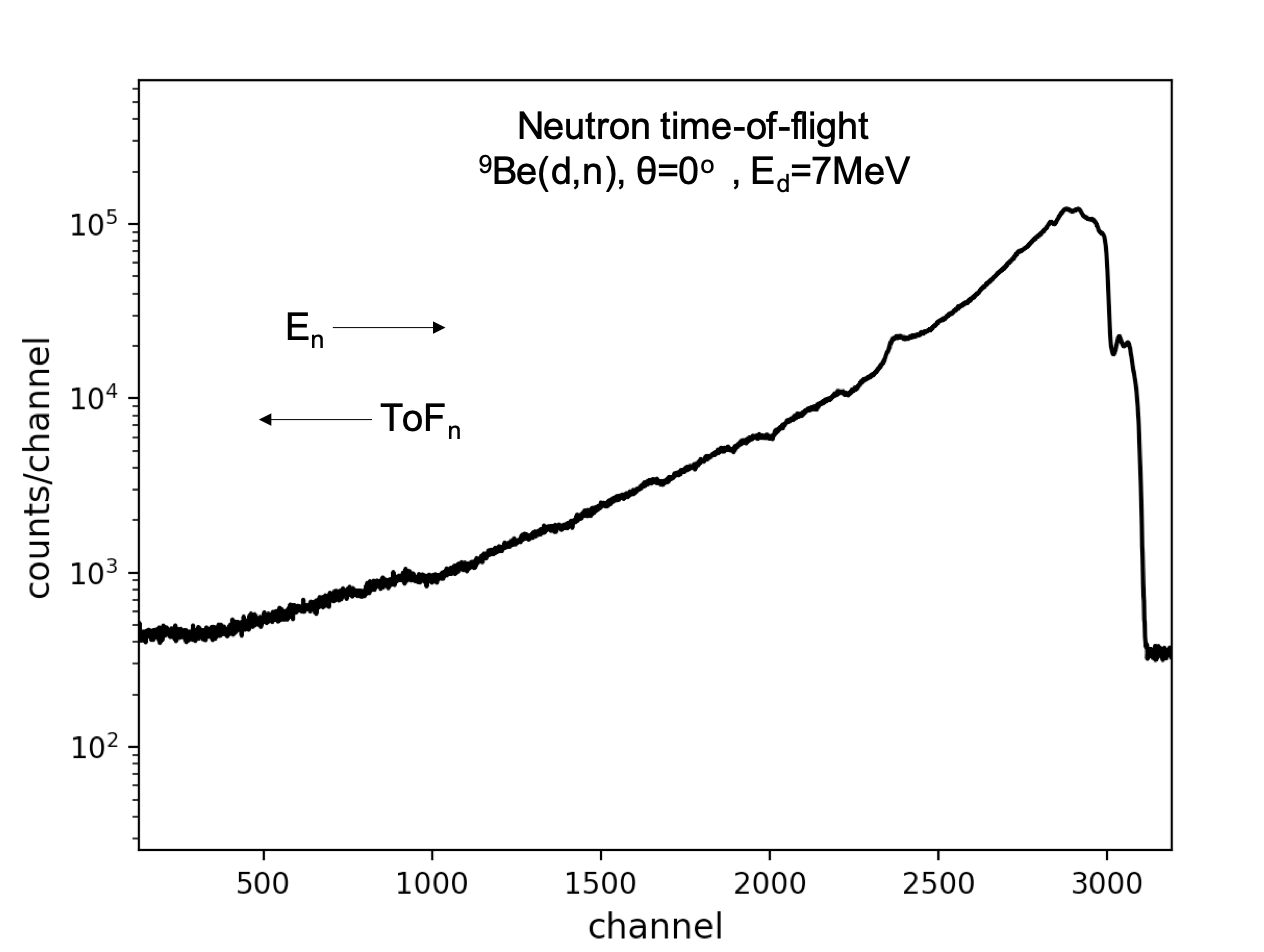}
\caption{Neutron time-of-flight spectrum obtained during the efficiency measurement. Neutron energy increases towards the higher channel numbers, while the corresponding time-of-flight increases in the opposite direction. }
\label{fig_tof_spectrum}
\end{figure}

The neutron time-of-flight spectrum of the $^{9}$Be(d,n) reaction that we obtained with the LENDA detector for the efficiency measurement is presented in Fig.~\ref{fig_tof_spectrum}. From the neutron time-of-flight, the kinetic energy of the neutrons was then deduced from:
\begin{equation}\label{eq_tof_to_e}
    E_{n} = \frac{m_{n}L^{2}}{2\tau^{2}c^2} 
\end{equation}
where $\tau$ is the neutron time-of-flight, $L$ is the flight path, $c$ is the speed of light, and $m_{n}$=939.57~MeV. 

\begin{figure}
\includegraphics[width=0.48\textwidth]{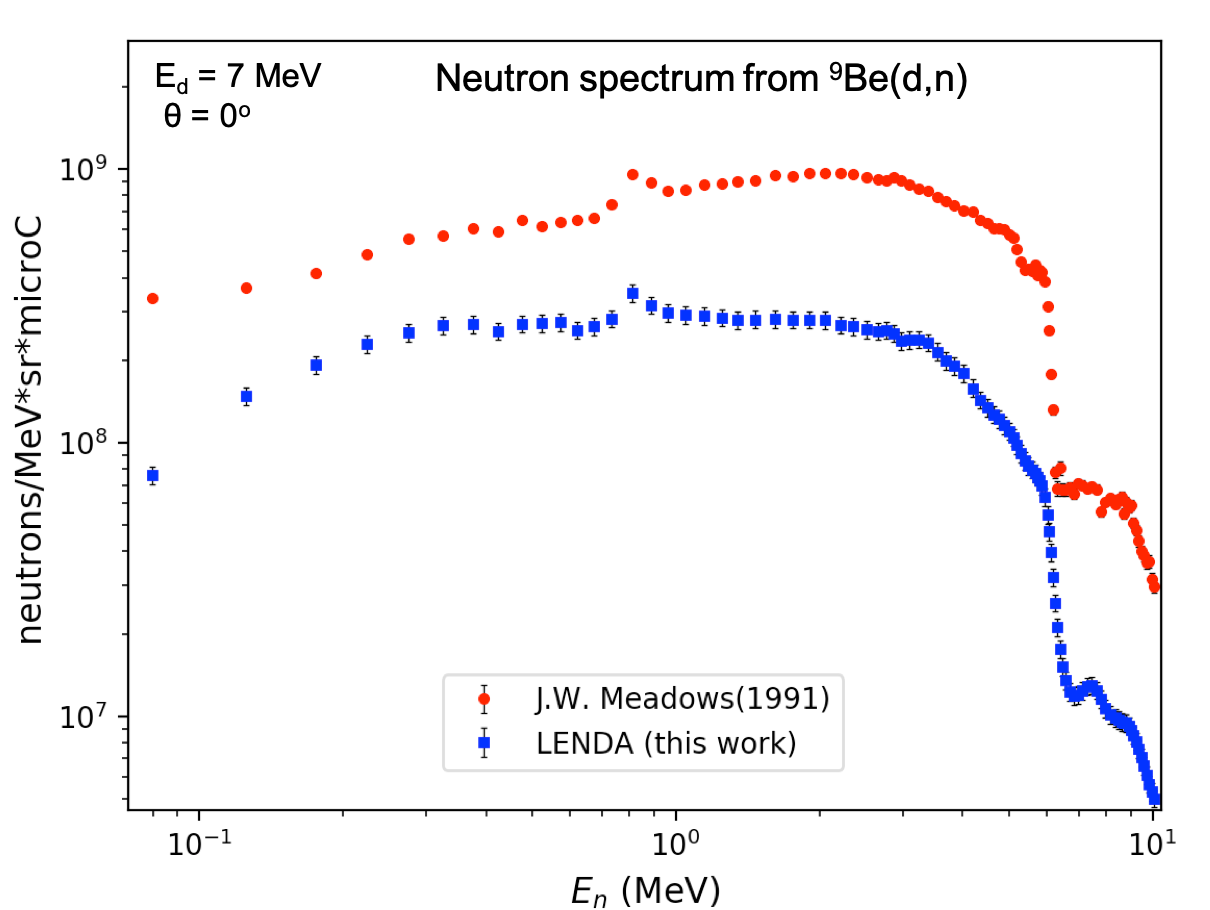}
\caption{Neutron yields from the $^{9}$Be(d,n) reaction. The reference spectrum has been adopted from \cite{Meadows:1991}. The intrinsic efficiency of the LENDA bar is reflected in the difference between the two curves. Each data point marks the center of the corresponding energy bin.}
\label{fig_yield}
\end{figure}

\begin{figure}
\includegraphics[width=0.48\textwidth]{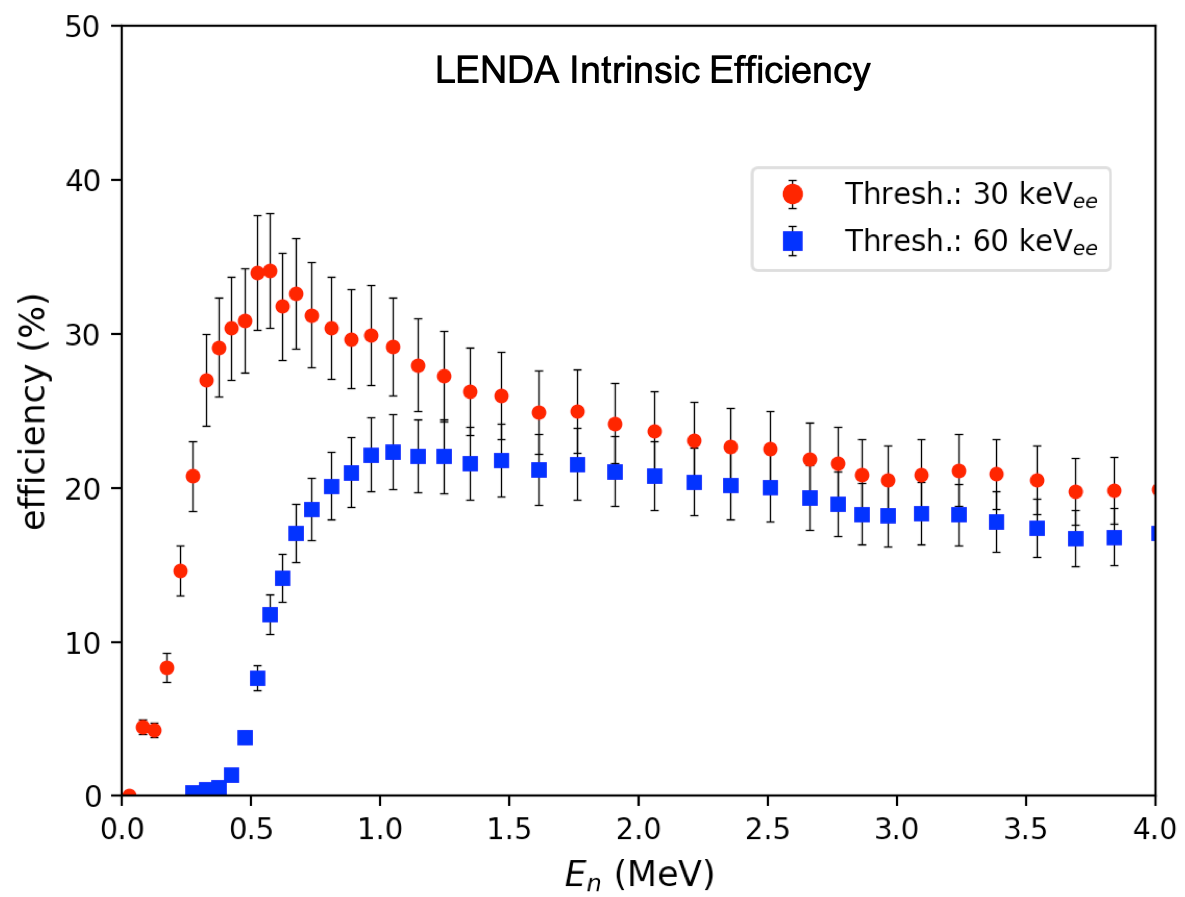}
\caption{Efficiency curve of the LENDA bar for two different threshold levels on the output pulse heights of the detector. The intrinsic efficiency below 1~MeV increases dramatically for the lower threshold and reaches a peak of approximately 35\% at 500~keV.}
\label{fig_lenda}
\end{figure}

We extracted the efficiency curve by comparing the measured neutron yields at each energy bin, with the ``standard" yields from this reaction. The standard yields were obtained from J.W. Meadows \cite{Meadows:1991}, while the measured yields were calculated as:
\begin{equation}\label{eq_yield}
    Y_{n} = \frac{ I_{n} }{\tau_{d} N_{b}\Delta\Omega \Delta E}
\end{equation}
where $I_n$ is the total number of counts in a particular bin, $\tau_{d}$ is a correction factor for the dead-time of the electronics system, $N_b$ is the integrated beam current (BCI) during the run, $\Delta\Omega$ is the solid angle of the detector, and $\Delta$E is the bin size. A comparison between the neutron yields from LENDA and those from the reference study is displayed in Fig.~\ref{fig_yield}. By adjusting the bin sizes accordingly, the efficiency at each energy bin was given from the ratio:
\begin{equation}\label{eq_efficiency}
    \epsilon_{n} = \frac{ [Y_{n}]_{LENDA}}{ [Y_{n}]_{Standard} }
\end{equation}
The extracted efficiency curve is presented in Fig.~\ref{fig_lenda}, where two different threshold levels were applied to the light output of the detector in offline analysis. For the analysis of neutron spectra from the $^{40}$Ar(p,n) reaction, both the efficiency curves were used. The 60~keV$_{ee}$ threshold was applied for beam energies above 3.6~MeV (center-of-mass system), and the 30~keV$_{ee}$ was applied for the lower energies to reflect the actual threshold of the detector during these measurements. %To minimize any unwanted beam-induced background, a threshold close to 60~keV$_{ee}$ was applied for beam energies above 4.0~MeV (lab system). On the other hand, for lower energies, the threshold reduced to $\sim$30~keV$_{ee}$ in order to detect the low energy neutrons from the $(p,n_2)$ channel.

The main uncertainties on the efficiency measurement were introduced by the beam current integration, the solid angle subtended by the detector, and the reference neutron yields. A detailed summary of the associated errors is given in Table~\ref{tab:errors_efficiency}.

\begin{table}[h]
\caption{Uncertainties on the various quantities of Eqs.~\ref{eq_yield} \& \ref{eq_efficiency}. Wherever a  range of values is used, it indicates that the corresponding error varied from data-point to data-point. The final uncertainty on the intrinsic neutron efficiency of LENDA was extracted by adding these errors in quadratic.}
\label{tab:errors_efficiency}
\begin{ruledtabular}
\begin{tabular}{cccc}
 & Quantity  & Uncertainty &  \\ \hline
 & Neutron yield (I$_n$) & $<$0.5~\%\footnotemark[1] &   \\
 & BCI (N$_b$) & 10.0~\%&   \\
 & Solid angle ($\Delta \Omega$) & 3.9~\%&   \\
 & Dead-time correction ($\tau_d$) & 0.2~\% & \\
 & Reference yield ([Y$_n$]$_{Standard}$) & 2.0 -- 5.0~\%\footnotemark[2] &   \\ \hline
 & Efficiency ($\epsilon_n$) & 10.9 -- 11.9~\% &   \\
\end{tabular}
\end{ruledtabular}
\footnotetext[1]{The error depended on the statistics of the corresponding energy bin.}
\footnotetext[2]{See~\cite{Meadows:1991}}
\end{table}

\subsubsection{\label{subsec_low_eff}Determination of efficiency for low energy neutrons} 

Due to the finite energy resolution of our system, which was dominated by the energy loss of the beam in the target, the widths of the neutron peaks from the $^{40}$Ar(p,n) reaction were about 100~keV. While the time-of-flight resolution was enough to separate the energies of neutrons coming from the target within 10~keV, all the neutrons within the 100~keV range would be included inside a single neutron peak. The energy resolution did not affect the analysis of the high energy neutron peaks (E$_n>$0.8~MeV) negatively as the efficiency is relatively constant within 100~kev at these energies (see Fig.~\ref{fig_lenda} for the efficiency at a threshold of 30~keV$_{ee}$). The sharp drop of LENDA's intrinsic efficiency for E$_n<$ 0.4~MeV, however, complicated the determination of absolute yields for the low energy neutron peaks (0.2 $<E_n<$ 0.4~MeV) in which the shape of the rapidly increasing efficiency was convoluted with that of the reaction yield. In other words, the energy width of each TOF-peak at low neutron energies corresponded to a significant variation of the neutron efficiency. We decoupled the effect of the varying efficiency inside the low energy neutron peaks by dividing them into intervals of 10-15~keV. For each one of these energy intervals, a weighting factor $w_i$ was obtained by dividing the number of counts in that interval over the total number of counts on the neutron peak. Having extracted a set of weighting factors, the cumulative efficiency of neutron detection for the events in each peak was calculated from the bin-wise yield as:
\begin{equation}\label{eq_wiei}
    \epsilon_{n} = \sum_{i=0}^{N} w_i \epsilon_i
\end{equation}
where N is the number of intervals, and $\epsilon_i$ is the intrinsic efficiency of LENDA at the corresponding energy of the $i^{th}$ interval.

%% ----------------------------------------
%%           GAMMA RAY SPECTRA
%% ----------------------------------------
\subsection{\label{sub_sec_gammarays}Gamma-ray spectra} 

\begin{figure}
\includegraphics[width=0.48\textwidth]{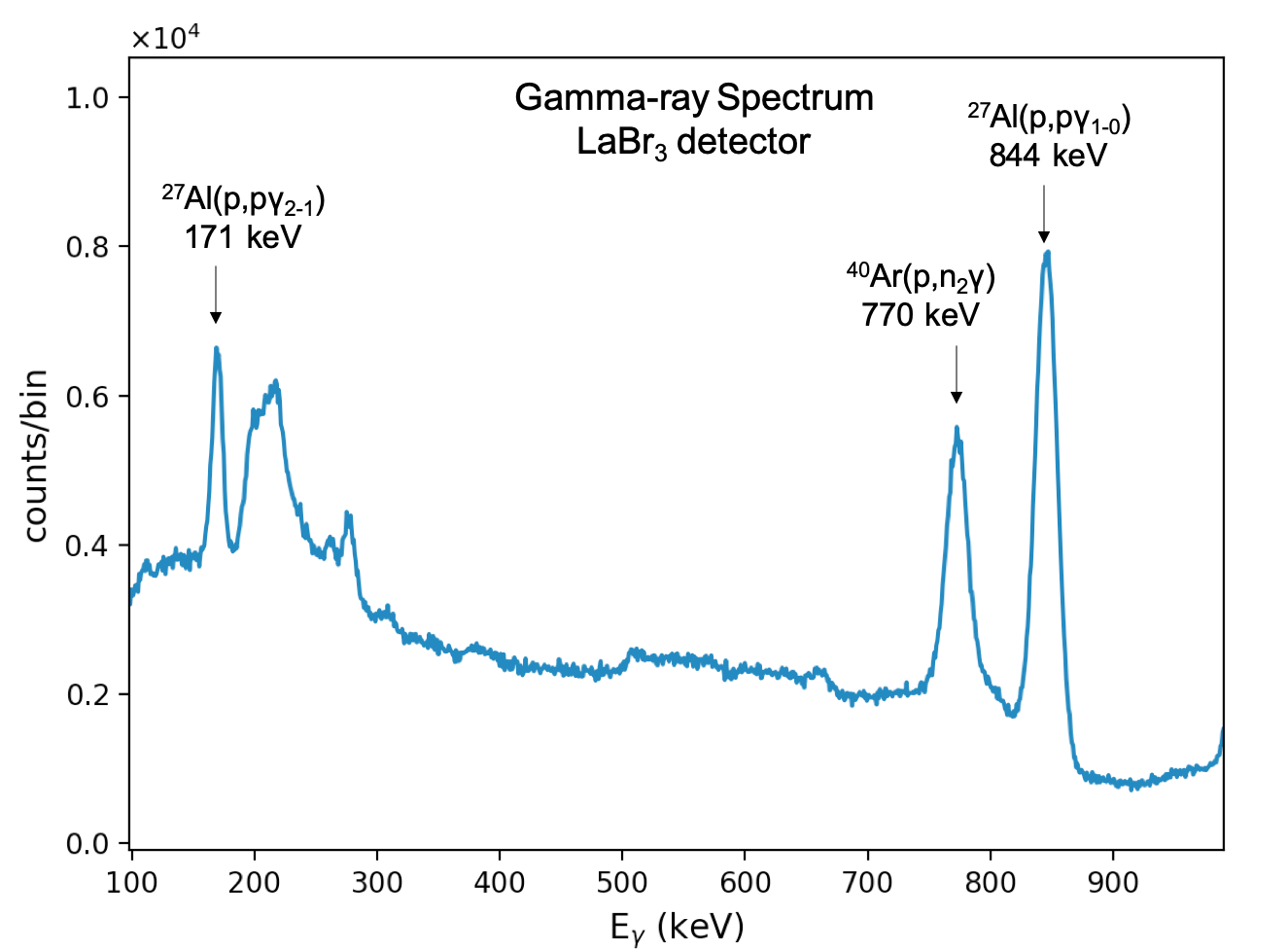}
\caption{Gamma-ray spectrum taken with the LaBr detector at E$_{lab}$ = 3.8~MeV. The marked photo-peaks are associated with the interactions of the proton beam with the aluminum windows of the gas-cell and the argon gas. Unlabeled peaks come from interactions of neutrons with the LaBr$_3$ crystal  the surrounding material in the experimental area, and from the detector's self-activity.}
\label{fig_gamma_spectrum}
\end{figure}

We used Gamma-ray spectra to study the characteristic gamma-rays from the $^{40}$Ar(p,n) reaction, as well as to cross-check the time-of-flight data. Fig.~\ref{fig_gamma_spectrum}, shows a gamma spectrum from the LaBr$_3$ detector for E$_{lab}$ = 3.8~MeV. The majority of the detected gamma-rays came from interactions of the proton beam with the argon gas, the aluminum window of the gas-cell, and the surrounding materials along the beamline. The 770~keV peak from the $^{40}$Ar(p,n$_2$) channel was present in the spectrum through all the measurements and was used to monitor the fluctuations of the cross-section. The 29.8~keV gamma-rays from the first excited state of $^{40}$K were also identified; however, they could not be easily resolved from the background photons in the low-energy region. In contrast with the first two excited states, the prominent gamma-rays from the $(p,n_{3})$ channel at 891~keV, were completely absent from the spectra. This was an indication that the partial cross-section of this channel is too low.

By determining the counting rates for the 770~keV gamma rays, we were able to investigate the behavior of the excitation function as a function of proton energy. This was done by extracting gamma-ray yields for the $(p,n_{2})$ channel at various incident energies. To avoid any systematic uncertainties associated with the dead-time of our electronics system, the yields were calculated relative to the 844~keV gammas-rays from the $^{27}$Al(p,p$\gamma_{1-0}$) reaction. Differential cross sections for the inelastic scattering on $^{27}$Al at 90-deg and incident energies between 3.0 and 4.0~MeV, were retrieved from M.Chiari et al. \cite{Chiari:2014} through the Ion Beam Analysis Nuclear Data Library (IBANDL) \cite{IBANDL:2019}. By using the 844~keV gamma-ray peak of aluminum as a reference, the relative yield factor Y$_{\gamma}$ of the $^{40}$Ar(p,n$_2-\gamma$) channel at each incident energy was calculated as:
\begin{equation}\label{eq_photo_rate}
    Y_{\gamma} = \left[\frac{d\sigma}{d\Omega}\right]_{Al} \frac{I_{\gamma}}{I_{Al}} A_{E}
\end{equation}
where $I_\gamma$ and $I_{Al}$ are the number of detected photons with energies 770~keV and 844~keV, respectively, [$d\sigma/d\Omega$]$_{Al}$ is the differential cross-section of the $^{27}$Al(p,p$\gamma_{1-0}$) reaction at the corresponding incident energy, and $A_{E}$ is a correction factor that takes into account the efficiency and the solid angle of the detector, as well as the areal density of $^{40}$Ar in the gas target.

%% ----------------------------------------
%%              RESULTS
%% ----------------------------------------
\section{\label{sec_results}Results and Discussion} 

\subsection{Qualitative interpretation of neutron and gamma spectra, and statistical fluctuations}

\begin{figure}[t]
\includegraphics[width=0.48\textwidth]{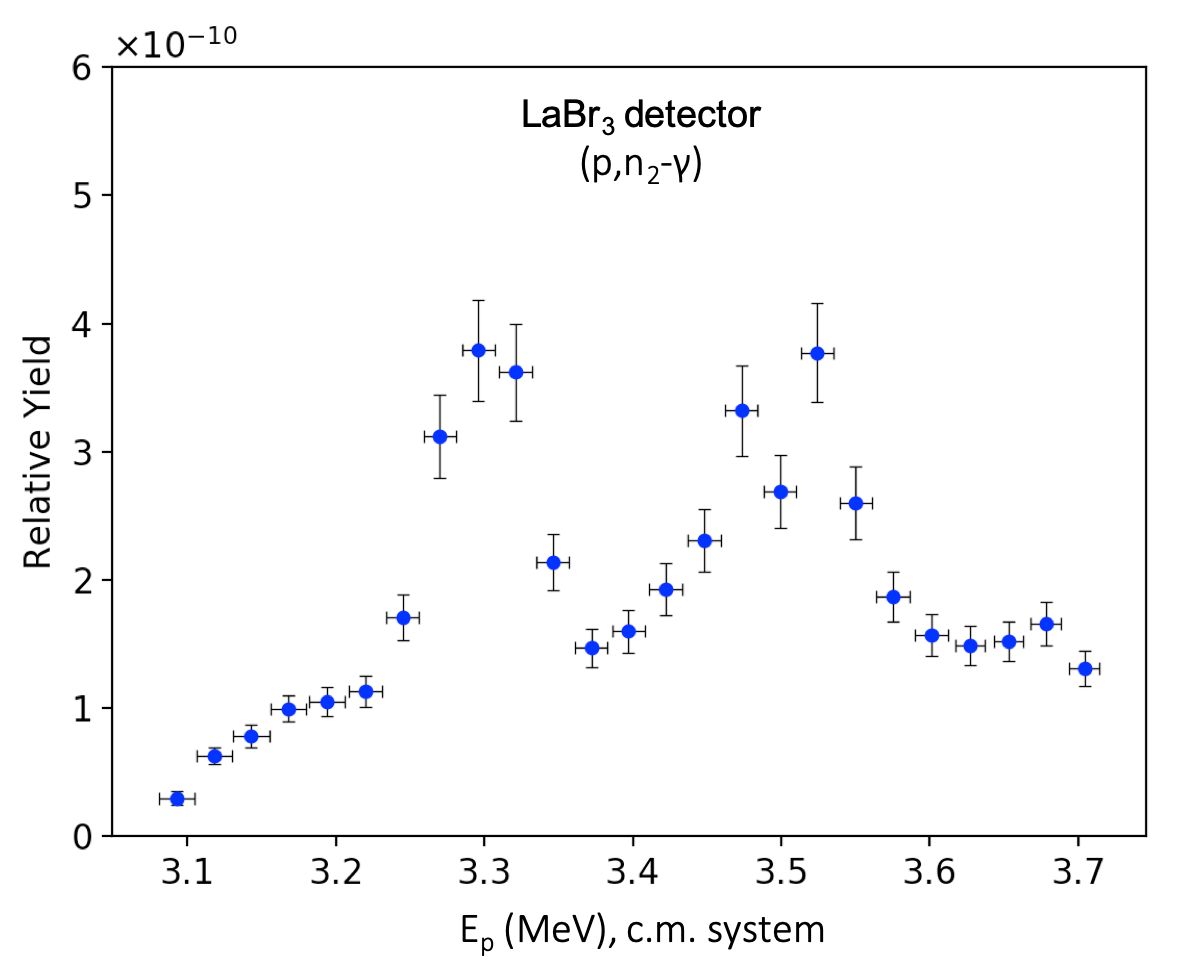}
\caption{Gamma-ray relative yields from the $(p,n_{2})$ channel. The two broad structures observed in the excitation function for incident energies above 3.2~MeV are attributed to two $^{41}$Ar Isobaric Analog State resonances folded with Ericson type statistical fluctuations of the cross-section. In the gamma spectra, the Ericson fluctuations are suppressed due to the 25~keV energy resolution of the measurement that makes them appear mostly as a slowly fluctuating ``background" continuum underneath the two broad IAS structures. The fluctuations were much more pronounced in the high-resolution time-of-flight spectra. There, the shape of the neutron peaks revealed a significant number of overlapping resonance-like structures (see discussion and Fig.~\ref{fig_neutron_energy_peaks}).}
\label{fig_sigma_LaBr}
\end{figure}

\begin{figure}[t]
\includegraphics[width=0.5\textwidth]{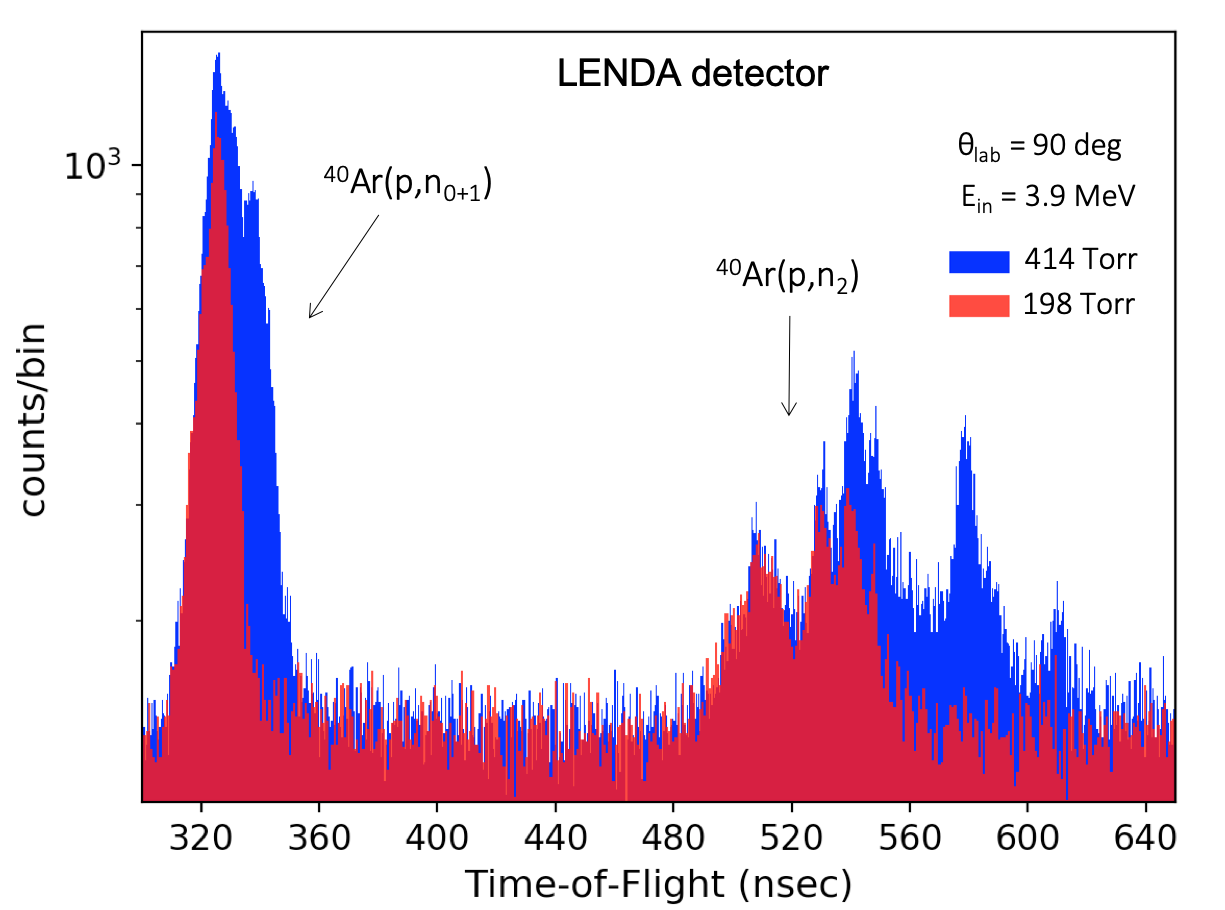}
\caption{Neutron time-of-flight spectra for two different target gas pressures, i.e., for two different target thicknesses. Both spectra correspond to the same beam energy (before interacting with the target window) and angle. The structured patterns on the neutron peaks are the result of the convolution of the Ericson fluctuations with the resonant neutron yields of the $^{41}$Ar isobaric analog states in $^{41}$K. For the blue spectrum, the reaction energy ranges between 3.4 and 3.6~MeV. For the red spectrum, the reaction energy ranges between 3.5 and 3.6~MeV. The difference in reaction energy range results in an integration of a larger portion of the fluctuating excitation function in the blue spectrum than in the red one. Consequently, more structures appear in the neutron peaks, consistent with what would be expected by a fluctuating cross-section (see discussion and Fig.~\ref{fig_sigma_LaBr}).}
\label{fig_neutron_energy_peaks}
\end{figure}

A remarkable feature of the experimental data was the fluctuating character of the cross-section. We observed fluctuations through both the yield of the 770~keV gamma-ray of the $(p,n_2)$ channel, and the neutron yield in the peaks of the time-of-flight spectra. The character of the fluctuations we observed in each case was dependent on the energy resolution of the corresponding measurement. Fig.~\ref{fig_sigma_LaBr}, shows the relative yield of the 770~keV gamma-ray as a function of beam energy. The 25~keV resolution of this phase of the measurement reveals a rapidly varying gamma-ray yield for reaction energies above 3.2~MeV in the center-of-mass, consistent with the known T=$5/2$ Isobaric Analog States (IAS) of $^{41}$Ar in $^{41}$K \cite{Scott1968a}. The IAS strength is mixed with the T=$3/2$ natural $^{41}$K states that have a much narrower width. This second contribution to the fluctuating character of the average cross-section is revealed in detail using the higher-resolution neutron time-of-flight spectra of Fig.~\ref{fig_neutron_energy_peaks}, that are taken with two different pressures (198 and 414~Torr) of the argon target gas. Both spectra correspond to the same incident energy and detector angle but have different argon target thicknesses. The proton energy of the beam after traversing the aluminum window of the gas target is 3.6~MeV. The energy resolution of the proton beam is 90 and 170~keV for the low and high-pressure case, respectively, and is dominated by the energy that the beam particles lose in the target gas before the onset of the nuclear reaction. Hence, in Fig.~\ref{fig_neutron_energy_peaks}, the blue spectrum corresponds to reaction energies in the range of 3.4--3.6~MeV, while the red spectrum corresponds to the range of 3.5--3.6~MeV. Both neutron spectra, therefore, correspond %roughly to the peak at 
the higher energy part of the gamma-ray spectrum of Fig.~\ref{fig_sigma_LaBr}.

For a non-fluctuating excitation function, the $(p,n_x)$ peaks in both spectra (Fig.~\ref{fig_neutron_energy_peaks}) would have the same single-peak shape with an energy spread defined by the folding of the IAS resonance width and the neutron energy resolution. The partial excitation functions, however,  are characterized by overlapping resonance-like structures, and the shape of the peaks changes according to the fraction of the excitation function covered in each measurement. The comparison of the two neutron spectra of Fig.~\ref{fig_neutron_energy_peaks} suggests that the shape of the neutron peaks is strongly affected by the folding of Ericson fluctuations with the IAS resonances, and the energy resolution of the measurement. As the beam energy loss increases in the target with the increased pressure, the portion of the excitation function probed increases too. As a result, the corresponding neutron peak becomes broader in energy (or time-of-flight), and the neutron yields reproduce a fluctuating excitation function for a broader energy range than before.

The energy resolution of the neutron peaks depends on the timing resolution of the system. Considering the kinetic energies of the neutrons and the geometry of our system, the neutron time of flight energy resolution during this experiment was of the order of 10~keV or less. For example, for the low energy neutron peaks that appear in Fig.~\ref{fig_neutron_energy_peaks} (right side), the resolution is $\sim$2.5~keV. Consequently, the statistical fluctuations of the cross-section are reproduced by the neutron yields in great detail and are seen in Fig.~\ref{fig_neutron_energy_peaks} superimposed on the IAS strength.

The fine structure we observed in the neutron spectra is typical of the region of nuclei we are studying and we attribute it to statistical fluctuations of the partial cross sections~\cite{Ericson:1960}. The observation of these structures is also in agreement with previous work in the literature using the $^{40}$Ar(p,n), (p,p), and (p,$\alpha$) reactions \cite{Scott1968a, Young1968a} that report on the superposition of the two types of contributions. Statistical fluctuations are expected in compound nuclear reactions that proceed via highly-excited overlapping states in the compound system and are in agreement with various other studies of nuclear reactions in this mass region where statistical fluctuations of the partial cross-sections were reported in literature~\cite{Witsch:1966,Naudi:1968,McMurray:1967}.

As discussed in~\cite{Ericson:1960}, the level interference effects corresponding to Ericson fluctuations impact both the partial cross-sections and the angular distributions. The partial angular distributions tend to be asymmetric relative to 90-deg, while the cross-sections tend to fluctuate as a function of energy with a period equal to the average width $\Gamma$ of the populated states. If the number of exit channels is small, the total cross-section is also affected. These effects are reduced when many exit channels are integrated or when the energy resolution of the measurement is much larger than $\Gamma$. As we will see in the following subsections, the results of our study agree with the predictions of the statistical fluctuations theory. The agreement is evident in the partial cross-sections and angular distributions, as well as in the result for the total cross-section. While the observed fluctuations are of particular interest, reporting a detailed fluctuation analysis of the results is complicated, particularly in the presence of the isobaric analog resonance states, and would be outside of the scope of this article. This analysis is, therefore left to be incorporated in future work. Here, we aim at the extraction of the reaction rate based on the measured energy averaged partial and total cross-sections.

%% ----------------------------------------
%%       TABLES - CROSS SECTIONS
%% ----------------------------------------

\subsection{Differential, partial, and total cross sections}

The measured partial cross sections for six energies are presented in Fig.~\ref{fig_TALYS_calc}. The  corresponding values are listed in Table~\ref{tab:results_sigma}. All energies correspond to the center-of-mass energy in the middle of the target, after taking into account the energy loss in the aluminum window and the argon gas. Energy losses were calculated using the Monte-Carlo software SRIM~\cite{SRIM}. 

A prominent drop of the cross-section is observed in all the neutron channels at around 3.6~MeV. This trend is consistent with the variation of the gamma-ray yields from the $(p,n_{2})$ channel (see Fig.~\ref{fig_sigma_LaBr} for $E_p>$ 3.4~MeV) in which Ericson fluctuations are seen folded into the broader T=5/2 isobaric analog resonance structure of $^{41}$Ar as discussed in the previous section. The 50~keV resolution of the total cross-section measurement results in a smoothing of the fluctuating character of the cross-section. The uncertainties on the partial cross-section determination fluctuated between 12.0 -- 13.5\% in $\sigma_{0,1}$ and 14.0 -- 22.5\% in $\sigma_{2}$. For the total cross-section $\sigma_{tot}$, the corresponding uncertainty range is 9.0 -- 11.5\%.

\begin{table}[h]
\caption{ Partial cross-sections of the $(p,n_{0,1})$ and $(p,n_{2})$ channels. The total cross-sections are given as $\sigma_{0,1}$ + $\sigma_{2}$    }
\label{tab:results_sigma}
\begin{ruledtabular}
\begin{tabular}{cccc}
 $E_{CM}$ (MeV)& $\sigma_{0,1}$ (mb)& $\sigma_{2}$ (mb) &$\sigma_{tot}$ (mb) \\ \hline
 3.882 $\pm$ 0.048 & 74.5 $\pm$ 9.6 & 26.1 $\pm$ 3.7 & 100.5 $\pm$ 10.5\\
 3.775 $\pm$ 0.049 & 71.9 $\pm$ 9.1 & 17.6 $\pm$ 2.5 & 89.5 $\pm$ 9.4\\
 3.687 $\pm$ 0.049 & 56.1 $\pm$ 6.8 & 9.6 $\pm$ 1.4 & 65.7 $\pm$ 7.0\\
 3.570 $\pm$ 0.050 & 91.3 $\pm$ 11.4 & 41.9 $\pm$ 5.9 & 133.2 $\pm$ 12.8\\
 3.463 $\pm$ 0.051 & 85.6 $\pm$ 11.5 & 37.7 $\pm$ 5.5 & 123.3 $\pm$ 12.7\\
 3.365 $\pm$ 0.051 & 53.1 $\pm$ 6.9 & 15.5 $\pm$ 3.4 & 68.6 $\pm$ 7.7 \\
\end{tabular}
\end{ruledtabular}
\end{table}

The extracted differential cross sections are given in Table~\ref{tab:results_diff_sigma}. Their uncertainties varied between 15.0-33.0\%. In general, uncertainties above 20.0\% mainly corresponded to measurements of low energy neutrons (E$_n\lesssim$ 0.4~MeV) from the $(p,n_{2})$ channel. This happened because the kinetic energies of those neutrons lie in a region where the intrinsic efficiency of LENDA changes rapidly (e.g., for E$_{cm}$=3.37~MeV, 200 $<E_n<$ 400~keV). Due to the convolution of the detector's efficiency with the neutron yields for those data points (see Section~\ref{sec_neutr_eff}), the total uncertainty was increased.

Overall, the primary sources of uncertainty were the beam current integration, the intrinsic neutron efficiency, the solid angle subtended by the neutron detector, and in some cases the peak integration. A detailed error budget is given in Table~\ref{tab:errors_diff}. Any systematic uncertainties associated with the shadowing of the target at 135-deg were estimated to be much smaller than any statistical error at this angle and were not treated separately.

\begin{figure}
\includegraphics[width=0.48\textwidth]{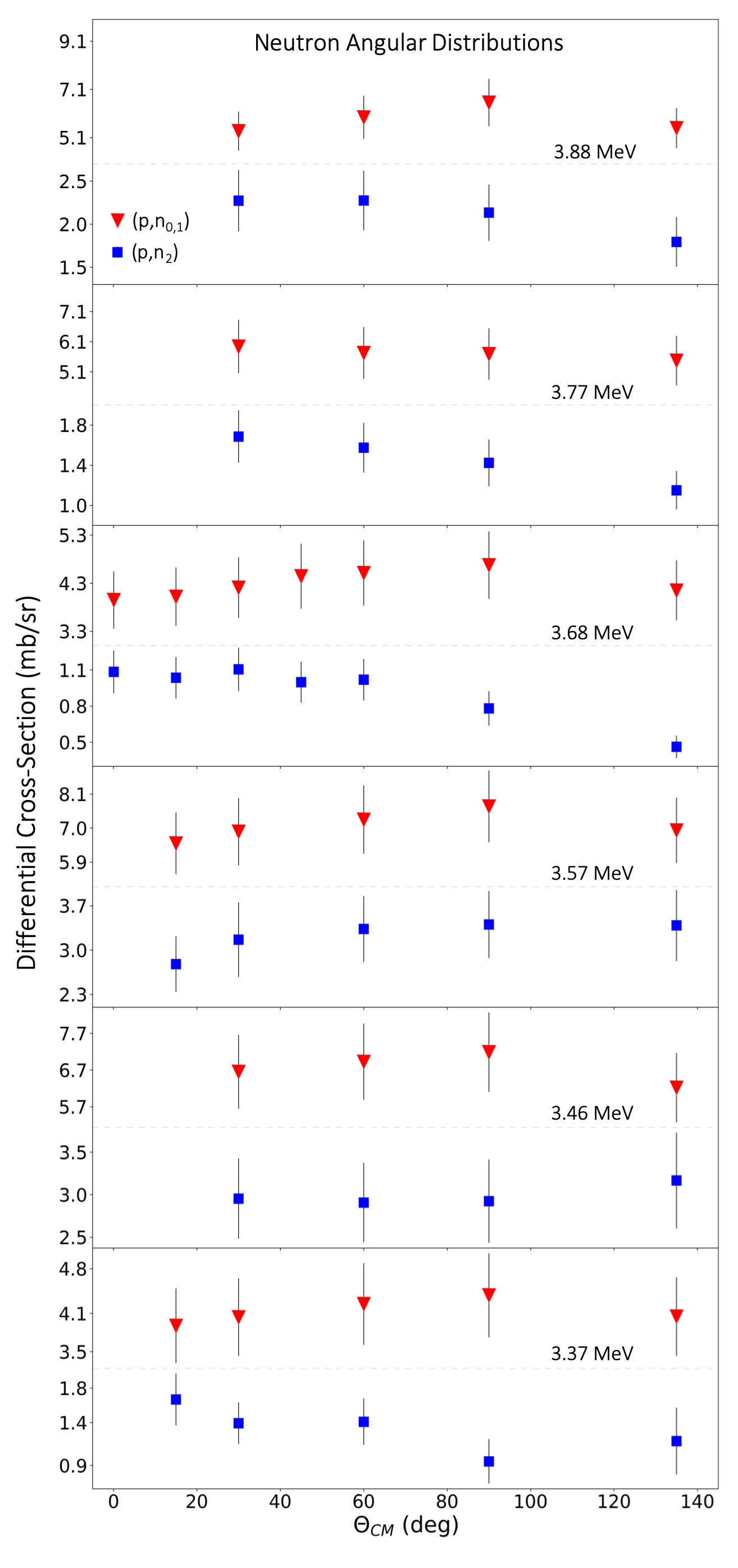}
\caption{Neutron angular distributions from $^{40}$Ar(p,n$_{0,1}$) and $^{40}$Ar(p,n$_{2}$) channels at various incident energies. All the energies and the differential cross-sections are given in the center-of-mass system.}
\label{fig_ang_distribution}
\end{figure}

In Fig.~\ref{fig_ang_distribution}, we present all the neutron angular distributions. The distributions of the $(p,n_2)$ channel were found to vary as a function of energy and tended to be asymmetrical relative to 90~$^\circ$. According to the theory of statistical fluctuations~\cite{Ericson:1960}, this behavior is expected for the distributions of individual channels. However, any features associated with this phenomenon should vanish when the distributions are averaged over a sufficiently large energy interval or when various exit channels are taken into account. The latter is demonstrated by the symmetrical distributions of the $(p,n_{0,1})$ channels, which change smoothly with energy. Furthermore, by looking at the energy averaged distributions (see Fig.~\ref{fig_av_calc_talys}) the profound asymmetries on the $(p,n_2)$ channel also vanish. This is an indication that the observed fluctuations of the cross-sections are purely statistical.

\begin{table}[h]
\caption{Uncertainties on the various quantities of Eq.~\ref{eq_dsigma}. Whenever a range of values is used, it indicates that the corresponding error varied from data-point to data-point. By adding these errors in quadratic, the final uncertainties on the differential cross-sections were extracted.}
\label{tab:errors_diff}
\begin{ruledtabular}
\begin{tabular}{cccc}
 & Quantity  & Uncertainty &  \\ \hline
 & Measured yield (I$_n$) & 1.0 -- 26.0~\%\footnotemark[1] &   \\
 & BCI (N$_p$) & 10.0~\%&   \\
 & Target thickness (N$_t$) & 5.0~\%&   \\
 & Solid angle ($\Delta \Omega$) & 3.9~\%&   \\
 & Dead-time correction ($\tau_d$) & 0.2~\% &   \\
 & Neutron efficiency ($\epsilon_n$) & 10.9 -- 15.0 ~\%\footnotemark[2] &   \\
\end{tabular}
\end{ruledtabular}
\footnotetext[1]{The error depended on the statistics of the corresponding neutron peak.}
\footnotetext[2]{The error varied for some neutron energies (see Section~\ref{subsec_low_eff}).  }
\end{table}

\begin{table*}
\caption{ Differential cross sections of the $^{40}$Ar(p,n$_{0,1}$) and $^{40}$Ar(p,n$_{2}$) channels. The error in angle $\theta$ is 0.25-deg. For keeping the table's format simple, all angles are given in the lab system.  }
\label{tab:results_diff_sigma}
\begin{ruledtabular}
\begin{tabular}{ccccccccc}

 \multicolumn{8}{c}{  }\\ 
 \multicolumn{8}{c}{  d$\sigma_{0,1}$/d$\Omega_{CM}$ (mb/sr)  }\\ 
   E$_{CM}$ (MeV) & 0 $^{\circ}$  
 & 15 $^{\circ}$ & 30 $^{\circ}$
 & 45 $^{\circ}$ & 60 $^{\circ}$
 & 90 $^{\circ}$ & 135 $^{\circ}$ \\ \hline
 
3.882 $\pm$ 0.048 & - & - & 5.4 $\pm$ 0.8  & - & 5.9 $\pm$ 0.9  & 6.5 $\pm$ 1.0  & 5.5 $\pm$ 0.8  \\
3.775 $\pm$ 0.049 & - & - & 5.9 $\pm$ 0.9  & - & 5.7 $\pm$ 0.9  & 5.7 $\pm$ 0.8  & 5.5 $\pm$ 0.8  \\
3.677 $\pm$ 0.049 & 4.0 $\pm$ 0.6  & 4.0 $\pm$ 0.6  & 4.2 $\pm$ 0.6  & 4.4 $\pm$ 0.7  & 4.5 $\pm$ 0.7  & 4.7 $\pm$ 0.7  & 4.1 $\pm$ 0.6  \\
3.570 $\pm$ 0.050 & - & 6.5 $\pm$ 1.0  & 6.9 $\pm$ 1.1  & - & 7.3 $\pm$ 1.1  & 7.7 $\pm$ 1.2  & 6.9 $\pm$ 1.0 \\
3.463 $\pm$ 0.051 & - & - & 6.7 $\pm$ 1.0  & - & 6.9 $\pm$ 1.0  & 7.2 $\pm$ 1.1  & 6.2 $\pm$ 0.9   \\
3.365 $\pm$ 0.051 & - & 3.9 $\pm$ 0.6  & 4.0 $\pm$ 0.6  & - & 4.2 $\pm$ 0.6  & 4.4 $\pm$ 0.7  & 4.0 $\pm$ 0.6 \\

\hline\\

 \multicolumn{8}{c}{  d$\sigma_{2}$/d$\Omega_{CM}$ (mb/sr) }\\ 
   E$_{CM}$ (MeV) & 0 $^{\circ}$  
 & 15 $^{\circ}$ & 30 $^{\circ}$
 & 45 $^{\circ}$ & 60 $^{\circ}$
 & 90 $^{\circ}$ & 135 $^{\circ}$ \\ \hline

3.882 $\pm$ 0.048 & - & - & 2.3 $\pm$ 0.4  & - & 2.3 $\pm$ 0.3  & 2.1 $\pm$ 0.3  & 1.8 $\pm$ 0.3   \\
3.775 $\pm$ 0.049 & - & - & 1.7 $\pm$ 0.3  & - & 1.6 $\pm$ 0.2  & 1.4 $\pm$ 0.2  & 1.1 $\pm$ 0.2  \\
3.677 $\pm$ 0.049 & 1.1 $\pm$ 0.2  & 1.0 $\pm$ 0.2  & 1.1 $\pm$ 0.2  & 1.0 $\pm$ 0.2  & 1.0 $\pm$ 0.2  & 0.8 $\pm$ 0.1  & 0.5 $\pm$ 0.1   \\
3.570 $\pm$ 0.050 & - & 2.8 $\pm$ 0.4  & 3.2 $\pm$ 0.6  & - & 3.3 $\pm$ 0.5  & 3.4 $\pm$ 0.5  & 3.4 $\pm$ 0.6  \\
3.463 $\pm$ 0.051 & - & - & 3.0 $\pm$ 0.5  & - & 2.9 $\pm$ 0.5  & 2.9 $\pm$ 0.5  & 3.2 $\pm$ 0.6   \\
3.365 $\pm$ 0.051 & - & 1.7 $\pm$ 0.3  & 1.4 $\pm$ 0.3  & - & 1.4 $\pm$ 0.3  & 0.9 $\pm$ 0.3  & 1.2 $\pm$ 0.4 \\

\hline\\

 \multicolumn{8}{c}{ d$\sigma_{tot}$/d$\Omega_{CM}$~(mb/sr)  }\\ 
   E$_{CM}$ (MeV) & 0 $^{\circ}$  
 & 15 $^{\circ}$ & 30 $^{\circ}$
 & 45 $^{\circ}$ & 60 $^{\circ}$
 & 90 $^{\circ}$ & 135 $^{\circ}$ \\ \hline
 
3.882 $\pm$ 0.048 & - & - & 7.6 $\pm$ 0.9  & - & 8.2 $\pm$ 0.9  & 8.7 $\pm$ 1.0  & 7.3 $\pm$ 0.9   \\

3.775 $\pm$ 0.049 & - & - & 7.6 $\pm$ 0.9  & - & 7.3 $\pm$ 0.9  & 7.1 $\pm$ 0.9  & 6.6 $\pm$ 0.8   \\
3.677 $\pm$ 0.049 & 5.0 $\pm$ 0.6  & 5.1 $\pm$ 0.6  & 5.3 $\pm$ 0.7  & 5.4 $\pm$ 0.7  & 5.5 $\pm$ 0.7  & 5.5 $\pm$ 0.7  & 4.6 $\pm$ 0.6  \\
3.570 $\pm$ 0.050 & - & 9.3 $\pm$ 1.1  & 10.0 $\pm$ 1.2  & - & 10.6 $\pm$ 1.2  & 11.1 $\pm$ 1.3  & 10.3 $\pm$ 1.2 \\
3.463 $\pm$ 0.051 & - & - & 9.6 $\pm$ 1.1  & - & 9.8 $\pm$ 1.1  & 10.1 $\pm$ 1.2  & 9.4 $\pm$ 1.1  \\
3.365 $\pm$ 0.051 & - & 5.6 $\pm$ 0.7  & 5.4 $\pm$ 0.6  & - & 5.7 $\pm$ 0.7  & 5.3 $\pm$ 0.7  & 5.2 $\pm$ 0.7  \\

\end{tabular}
\end{ruledtabular}
\end{table*}

\subsection{Theoretical calculations}\label{sec_talys_results}
To reproduce the experimental data, we performed statistical model calculations for incident energies between 3.0 and 4.0~MeV in the center-of-mass system, by using the Hauser-Feshbach code TALYS 1.9~\cite{talys1.0}. In these calculations, we adopted the semi-microscopic Optical Model Potential (OMP) of Jeukenne-Lejeune-Mahaux \cite{Bauge:2001rr}, while for the description of level densities and gamma-ray strength functions we used the Back-shifted Fermi gas model~\cite{Dilg:1973zz}, and the Brink-Axel model~\cite{Brink:1957,Axel:1962zz} respectively. We obtained no significant variation of the calculation results when we tried different models for level densities and gamma-strength functions. Regarding the nuclear masses involved, we have implemented the most recently evaluated atomic mass data \cite{masses_2017}.

The three exit channels which are competing in the p + $^{40}$Ar reaction, are the (p,p), (p,n), and (p,$\alpha$). The parameters of the optical model for each one of the three possible particles in the exit channel (neutrons, protons, and alphas) are expected to influence the result of the cross-section calculation. To get a better description of the partial cross-sections for the (p,n) channel, we adjusted the OMP transmission coefficients only for neutrons and alphas. The alpha-nucleus optical potential is typically more uncertain than the proton-nucleus one (see e.g., \cite{Perreira2016a}). Therefore, to limit the number of tuned parameters, the corresponding proton transmission coefficients remained unchanged.
The transmission coefficients were adjusted using the keyword `Tljadjust' of TALYS, while all the other parameters in the code remained fixed. The `Tljadjust' parameter allows the user to multiply the transmission coefficients with a different factor for each particle and orbital angular momentum L. The range of L-values for which we calculated the transmission coefficients depends on the populated spins in the compound system. We applied corrections up to the maximum L that was given by the models used in the TALYS calculation for the energy range of our interest. The limit of L$_{max}$ was 5 for neutrons, and 11 for alphas.
The optimum set of multiplication factors for the transmission coefficients was extracted based on the quality of the cross-section fits. For the different input parameters, the fit quality was evaluated by using Pearson's chi-squared formula:
\begin{equation} \label{chi_sq}
    \chi ^{2}  = \sum_{j=1}^{N} \frac{ (E_{j} - O_{j})^{2} } {O_{j} }  
\end{equation}
where $E_{j}$ and $O_{j}$ are the experimental and theoretical cross-sections respectively.

\subsubsection{Calculations of Cross-Sections and Angular distributions}\label{subsec_talys_results_A}

\begin{figure}
\includegraphics[width=0.48\textwidth]{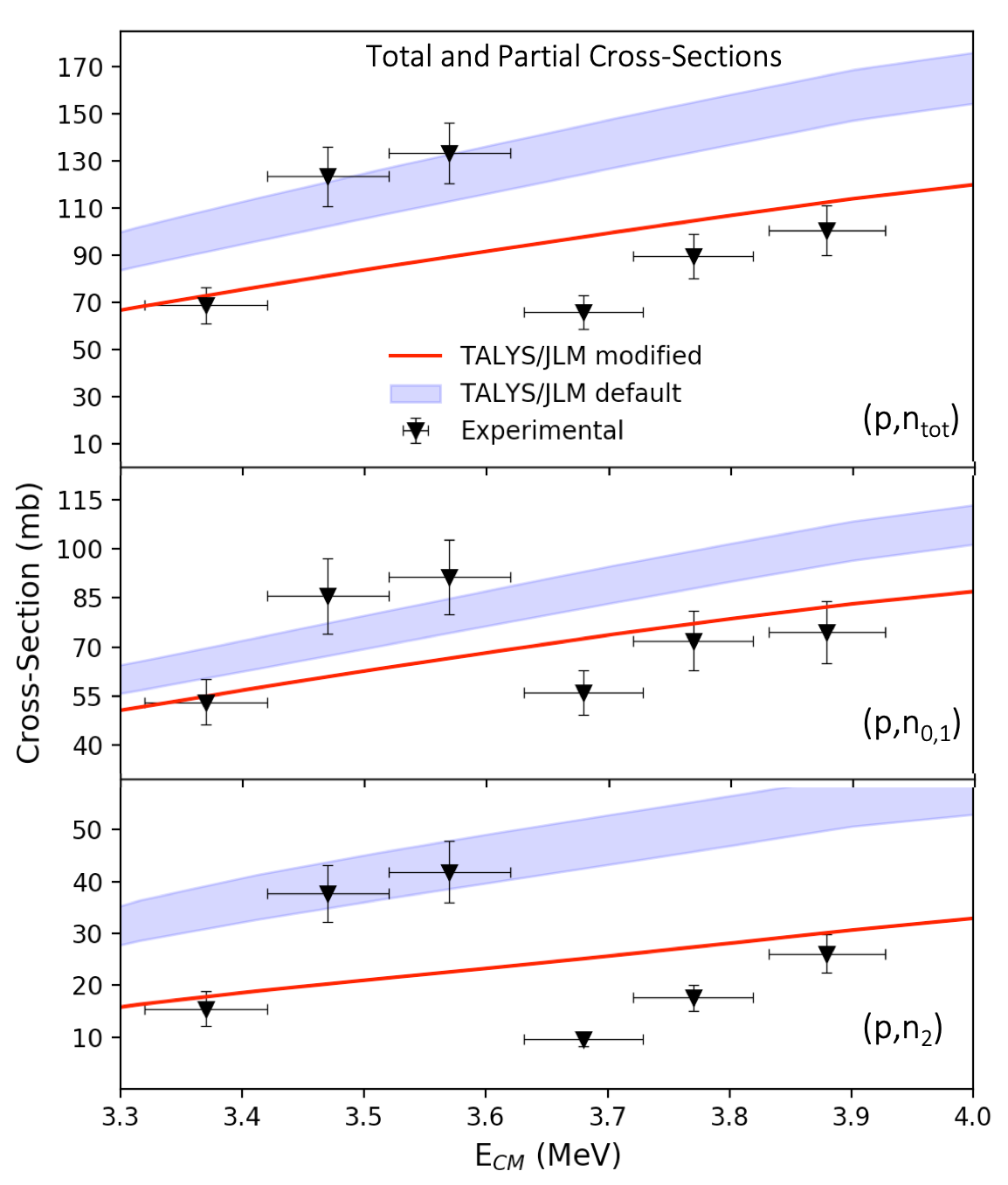}
\caption{Experimental and theoretical cross-sections for the $^{40}$Ar(p,n)$^{40}$K reaction. The theoretical cross-sections have been calculated using the TALYS code as described in the text. The error-band shows the maximum variations of the calculations due to uncertainties in the tabulated JLM optical potential parameters include with TALYS (default label). The ``modified" model, is a calculation in which the neutron and alpha transmission coefficients using the same tabulated JLM parameters \cite{Bauge:2001rr}, were adjusted to properly reproduce the experimental data. }
\label{fig_TALYS_calc}
\end{figure}

\begin{figure}
\includegraphics[width=0.48\textwidth]{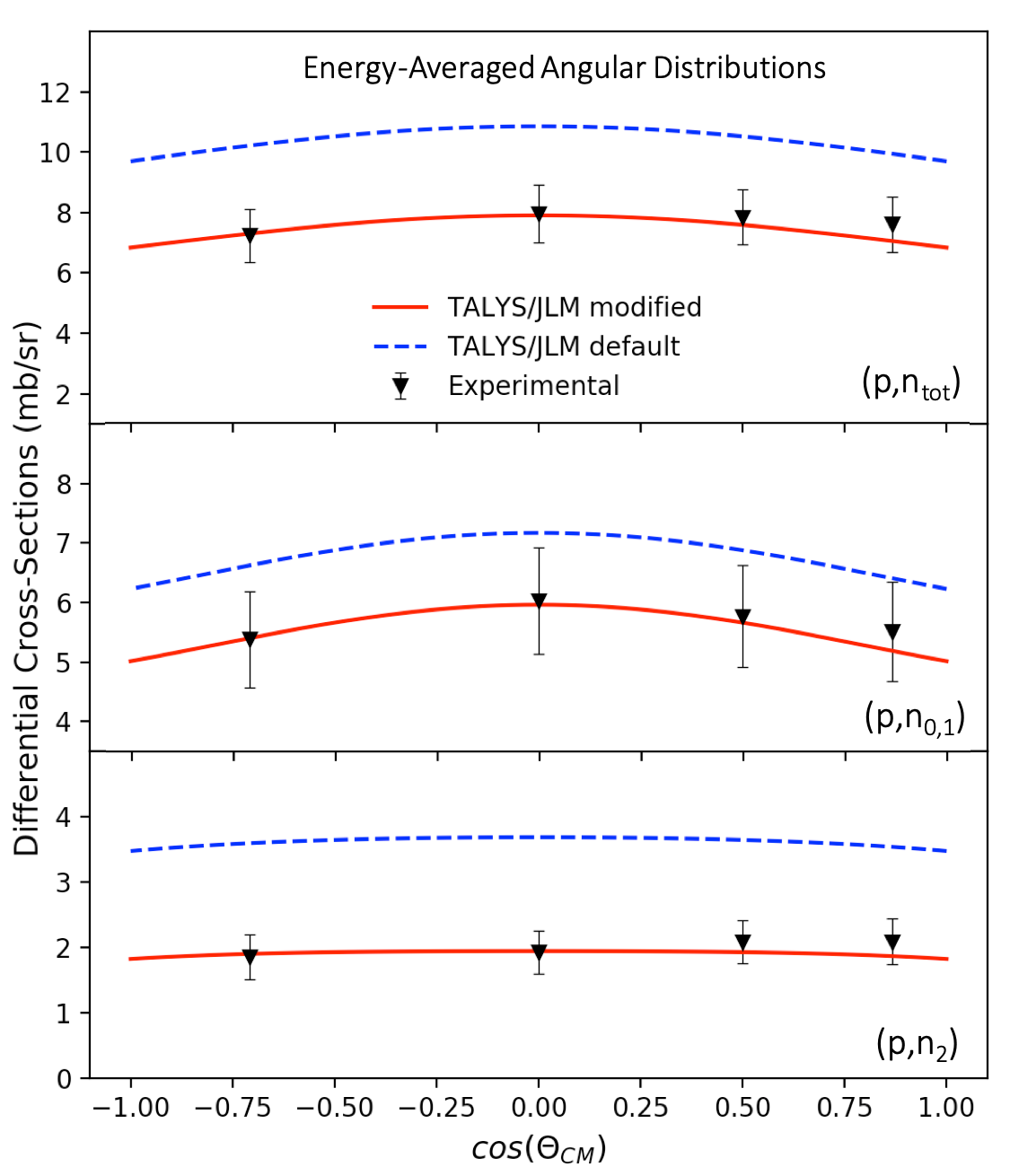}
\caption{Experimental and theoretical energy-averaged angular distributions of the $^{40}$Ar(p,n)$^{40}$K reaction. The distributions are averaged over the energy range 3.33 to 3.92~MeV. A better agreement between the results, is observed when the modified transmission coefficients are used (see also Fig.~\ref{fig_TALYS_calc}).}
\label{fig_av_calc_talys}
\end{figure}

The results from the statistical model calculations are summarized in Figs.~\ref{fig_TALYS_calc} \& \ref{fig_av_calc_talys}. Since the cross-sections were characterized by significant fluctuations, the data cannot be precisely reproduced within the Hauser-Feshbach framework that calculates energy-averaged cross-sections. We expect, however, that the energy-averaged angular distributions will be reproduced accurately in shape and magnitude, and the general trend of the cross-sections variation as a function of energy should be reproduced reasonably. Here we make a comparison between theoretical and experimental cross-sections. We evaluate the agreement of the calculation with data using the chi-square deviation of the two (see Eq.~\ref{chi_sq}). 

By using the default -tabulated- parameters for the real and imaginary OMP components in TALYS, all the corresponding partial cross-sections were overestimated. To investigate the impact of the existing uncertainties on these parameters, we repeated the calculations after varying the JLM normalization factors $\lambda_V$ (real central), $\lambda_{V1}$ (real isovector), $\lambda_{W}$ (imaginary central), and $\lambda_{W1}$ (imaginary isovector) within their suggested limits \cite{Bauge:2001rr}. The maximum variations of the theoretical cross-sections are presented as error-bands in Fig.~\ref{fig_TALYS_calc}. On average, the deviations from the best fit-line were around 20\% for (p,n$_{0,1}$), and 100\% (factor of two) for (p,n$_{2}$). The latter disagreement amplifies the discrepancy between theory and experiment for the total cross-section, which was of the order of 30\%.

For getting a more accurate description of the partial cross-sections within the statistical model approach, we modified the initial alpha and neutron transmission coefficients as described at the beginning of this section. Specifically, we increased all the alpha transmission coefficients by a factor of 2.2, while for neutrons, we decreased them by 60\% for L=0,1 and 10\% for L$>$1. The final transmission coefficients are given in Tables \ref{results_transmission_coeff_neut},\ref{results_transmission_coeff_prot},\ref{results_transmission_coeff_alp}.

The partial cross-sections of the (p,n$_{0,1}$) channels were found to be equally sensitive to all the involved transmission coefficients. On the other hand, the (p,n$_{2}$) channel was mainly sensitive to the L=0 and L=1 components. For this reason, the main modifications to the neutron transmission coefficients were done for the low L-values. This can be explained by the fact that in our models, the populated spin distribution in the compound system peaks at J=2.5. Since the 2nd excited state of the residual nucleus ($^{40}$K) has J=2, an outgoing neutron from this transition is more likely to have L=0 or L=1.

By using the modified transmission coefficients, we calculated the neutron angular distributions for the various exit channels. The results are presented in Fig.~\ref{fig_av_calc_talys}, where the calculated distributions are compared with the experimental data. To minimize the effects of the statistical fluctuations, we compared the energy averaged distributions. The average was evaluated over the range 3.33 -- 3.92~MeV in the center-of-mass system. 
The theoretical angular distributions were found to be in a better agreement with the experimental data when the modified transmission coefficients were used. This result is consistent with the calculations on the partial and total cross sections and verifies the aptness of our modifications. The fact that a significant part of the observed discrepancies on the total cross sections was related to the (p,n$_{2}$) channel, highlights the importance of analyzing individual exit channels when constraining nuclear models.

\subsubsection{\label{subsec_talys_results_B}Astrophysical reaction rates}

\begin{figure}
\includegraphics[width=0.49\textwidth]{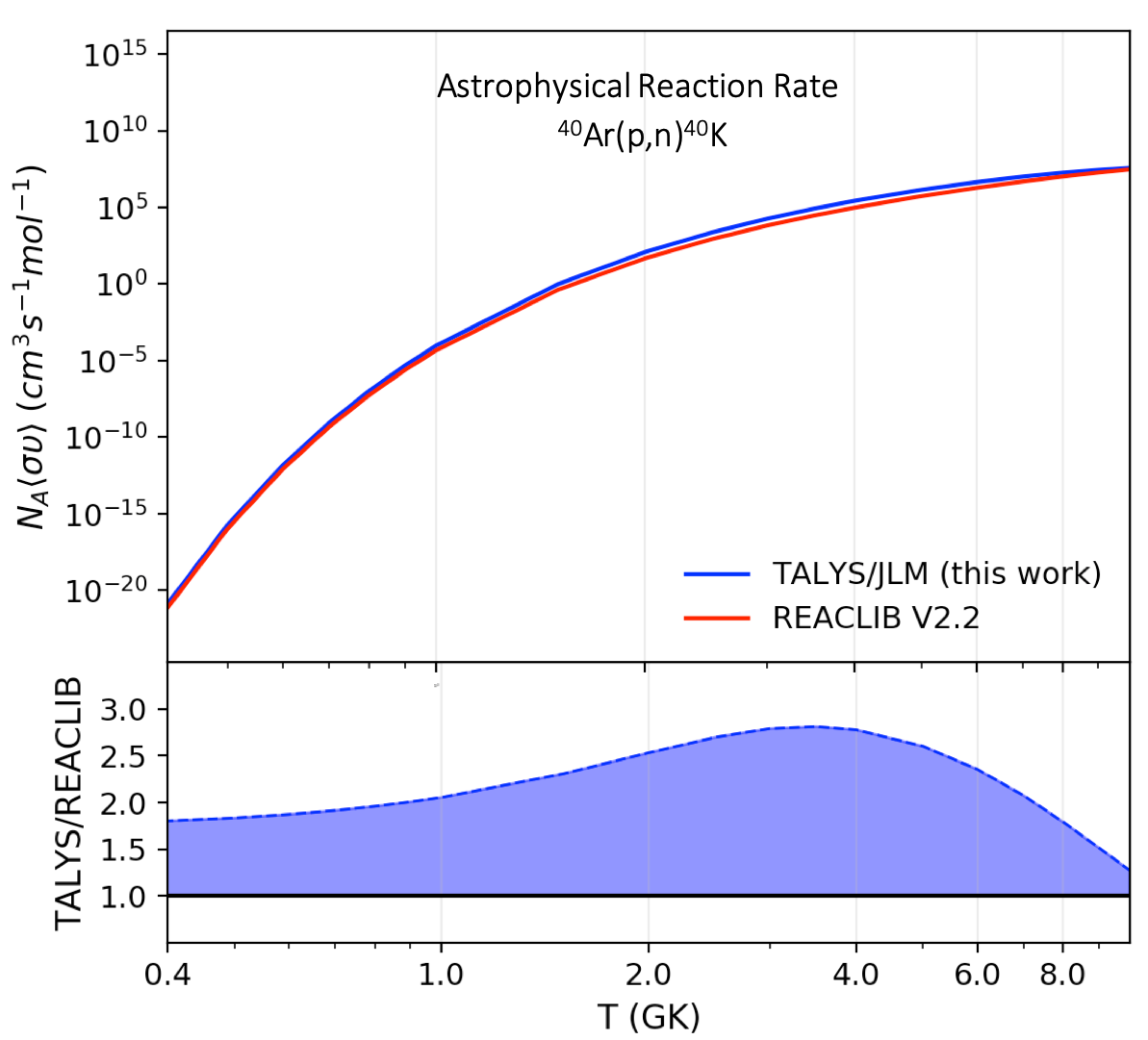}
\caption{(\textbf{Top}) Calculated astrophysical reaction rate of the $^{40}$Ar(p,n)$^{40}$K reaction, using the modified transmission coefficients from this work. The corresponding reaction rate of REACLIB database is also included (``rath" rate by Rausher et al. \cite{Rauscher_2000}). (\textbf{Bottom}) Comparison between the reaction rates from this work and the REACLIB database. The vertical axis corresponds to the ratio of $Rate_{TALYS}/ Rate_{REACLIB}$.}
\label{fig_astro_rate_pn}
\end{figure}

In Fig.~\ref{fig_astro_rate_pn}, we present the calculated astrophysical reaction rate for the $^{40}$Ar(p,n)$^{40}$K reaction, using the modified transmission coefficients from this study. The calculation was performed in TALYS using the keyword ``astro y". In this mode, TALYS calculates the reaction rate for the target in the ground state as well as in a distribution of excited states populated according to a Boltzmann distribution. The detailed results are given in Table~\ref{tab:reaction_rates}. The current calculation of the thermonuclear rate is based on the experimental cross-section that we measured in this work using proton energies between 3 and 4~MeV. The equivalent neutron energies for the reverse reaction that destroys $^{40}$K, i.e., the exothermic $^{40}$K(n,p)$^{40}$Ar reaction, would lie between 1.0 and 1.6~MeV. Considering the Maxwell-Boltzmann distribution widths (and the Gamow-Window for protons), the energies that we performed the measurement correspond to astrophysical temperatures above 4.5~GK. 
By comparing the experimentally constrained rate with the one of REACLIB (see Fig.~\ref{fig_astro_rate_pn}), we observe that above 4.5~GK the two rates differ by over a factor of two. In the range below 1~GK, which is relevant to stellar nucleosynthesis, the difference is over 80\%. 

To extract the reaction rate of the $^{40}$K(n,p)$^{40}$Ar reaction, we used the detailed balance condition \cite{Rauscher_2000}:
\begin{eqnarray}
 \label{eq_reac_rate}
N_A\langle \sigma \upsilon \rangle_{(n,p)} & & = 
\left( \frac{A_{Ar} A_{p}}{A_{n} A_{K}} \right)^{\frac{3}{2}}  \frac{(2J_{Ar}+1) (2J_p +1) }{(2J_K +1) (2J_n +1)}\nonumber\\
& & \times \frac{G_{Ar}(T)}{G_{K}(T)}
\times e^{ -Q_{(p,n)}/kT} \times N_A\langle \sigma \upsilon \rangle_{(p,n)}
\end{eqnarray}
where $N_A\langle \sigma \upsilon \rangle$ are the corresponding reaction rate for the forward and reverse reactions ($N_A$ is the Avogadro number), $A$ and $J$ are the mass numbers in a.m.u and the ground state spins of the involved nuclei, $k$ is the Boltzmann constant, T is the temperature, $Q_{(p,n)}$ is the Q-value of the endothermic reaction, and $G(T)$ are the temperature-dependent partition functions that correspond to the excited state spectrum of the Ar and K nuclei (calculated using the TALYS code). Using the above formula both the ground and excited states in the entrance and exit channel of the reaction are taken into account in the calculation of the reverse reaction rate. The results of these calculations are given in Table~\ref{tab:reaction_rates}, along with the partition functions that we used.

\begin{figure}
\centering
\includegraphics[width=0.49\textwidth]{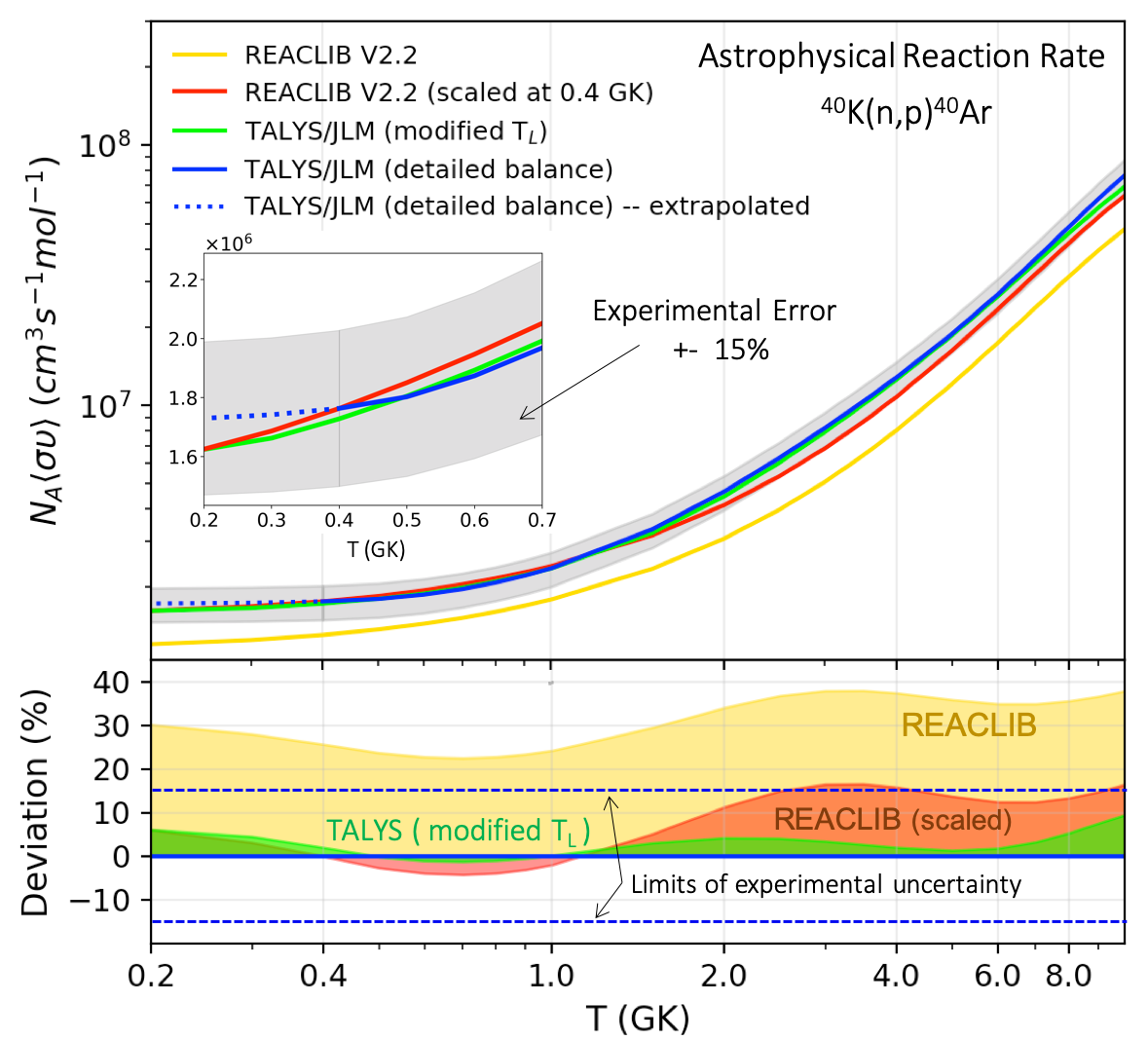}
\caption{(\textbf{Top}) Constrained astrophysical reaction rate of the $^{40}$K(n,p)$^{40}$Ar reaction (\textit{blue line}). The rate is deduced from the Hauser-Feshbach calculations on the $^{40}$Ar(p,n)$^{40}$K reaction, using the principle of detailed balance (see Eq.~\ref{eq_reac_rate}). For temperatures below 0.4~GK an exponential extrapolation was applied as described in the text. The grey error-band, marks the limits of the average experimental uncertainty. The \textit{green line}, represents the same reaction rate but calculated using the modified transmission coefficients from this study (Hauser-Feshbach calculation with TALYS). In the same graph, the reaction rate from the REACLIB library is also displayed (``rath" rate by Rausher et al. \cite{Rauscher_2000}, \textit{yellow line}). The \textit{red line}, corresponds to the REACLIB rate after being scaled to a reference value at 0.4~GK. 
(\textbf{Bottom}) Comparison between the various reaction rates, relative to the reference rate obtained from detailed balance. The deviations are calculated using Eq.~\ref{eq_deviation}. The REACLIB rate is systematically above the limits of the experimental uncertainty, and deviations up to $\sim$40\% are observed. On the other hand, the scaled REACLIB rate and the TALYS rate (with modified T$_L$) are in a very good agreement with the reference curve. }
\label{fig_astro_rate_np}
\end{figure}

In Fig.~\ref{fig_astro_rate_np}, we compare the extracted rate for the $^{40}$K(n,p)$^{40}$Ar reaction (blue line) with the recommended values adopted in the REACLIB V2.2 library (yellow line). %The extrapolated data are displayed with the dashed line for temperatures between 0.2 and 0.4~GK. 
In the same graph, we include the corresponding reaction rate calculated using the modified transmission coefficients from this study (green line). A detailed comparison of the reaction rates is shown in the lower panel of Fig.~\ref{fig_astro_rate_np}, where the deviation D is defined as:
\begin{eqnarray}
 \label{eq_deviation}
D = 100\times\frac{R_{DB} - R_i}{R_{DB}}
\end{eqnarray}
where, R$_{DB}$ is the reaction rate obtained from detailed balance, and R$_i$ is the calculated rate using any other method. The calculated rate of the (p,n) reaction gave zero values for T$<$0.4~GK preventing the calculation of the detailed-balance (n,p) below that temperature. To extend the detailed-balance data-set to temperatures below 0.4~GK, which are relevant for stellar nucleosynthesis, we chose to apply an exponential extrapolation (dashed blue line). The extrapolation was done based on the calculated rates at the temperature range between 0.4 and 1~GK. This extrapolation retained the trend of the detailed-balance rate below 0.4~GK without producing a nonphysical kink as it would be the case if we used one of the modified TALYS or scaled REACLIB calculations to extrapolate. We consider this extrapolation choice justified, since in the range between 0.2 -- 8~GK, the reaction rates from detailed balance (solid and dashed blue), scaled REACLIB (red), and TALYS (green) are in agreement within 5\%. Moreover, this discrepancy is well within the average experimental uncertainty of $\sim$15\% (grey error-band), while the smoothness of the extrapolation is retained. In contrast to this agreement, the REACLIB rate differs by -up to- 40\% from the experimentally constrained ones in the temperature range between 0.2 -- 10~GK. This discrepancy is expected to have implications on stellar nucleosynthesis calculations for the abundance of $^{40}$K produced by the s-process, as the $^{40}$K(n,p)$^{40}$Ar reaction is partially responsible for its destruction.

\begin{table*}
\caption{ Experimentally constrained reaction rates based on the cross-sections of this study. For the (p,n) reaction, all the rates below 0.4~GK are zero. The reverse (n,p) rate was calculated by applying the detailed balance conditions (see Section~\ref{subsec_talys_results_B}). For temperatures below 0.4~GK, the reverse rate was extracted by doing exponential extrapolation (see Fig.~\ref{fig_astro_rate_np}). Taking into account the experimental errors, the tabulated reaction rates are given within an uncertainty of 15\%. The corresponding partition functions for thermalized $^{40}$Ar and $^{40}$K targets were calculated in TALYS.}
\label{tab:reaction_rates}
\begin{ruledtabular}
\begin{tabular}{ccccc}
   &  \multicolumn{2}{c}{Reaction Rate ($cm^3s^{-1}mol^{-1}$)} &  \multicolumn{2}{c}{Partition Functions}  \\
 \cline{2-3} \cline{4-5} 
 \rule{0pt}{0.35cm} T (GK)  & $^{40}$Ar(p,n)$^{40}$K & $^{40}$K(n,p)$^{40}$Ar &  $G_{Ar}$(T*)  &  $G_{K}$(T*) 
 \\ \hline
 
0.2 & 0.000000E+00  &  (1.729863E+06)  &  1.00000E+00  &  1.13777E+00 \\
0.3 & 0.000000E+00  &  (1.741773E+06)  &  1.00000E+00  &  1.24531E+00 \\
0.4 & 3.253910E-22  &  1.763392E+06  &  1.00000E+00  &  1.32734E+00 \\
0.5 & 2.014160E-16  &  1.802633E+06  &  1.00000E+00  &  1.38920E+00 \\
0.6 & 1.503560E-12  &  1.873862E+06  &  1.00000E+00  &  1.43680E+00 \\
0.7 & 8.989150E-10  &  1.968621E+06  &  1.00000E+00  &  1.47433E+00 \\
0.8 & 1.109410E-07  &  2.083213E+06  &  1.00000E+00  &  1.50458E+00 \\
0.9 & 4.782250E-06  &  2.215639E+06  &  1.00000E+00  &  1.52946E+00 \\
1.0 & 9.869730E-05  &  2.364713E+06  &  1.00000E+00  &  1.55028E+00 \\
1.5 & 1.010580E+00  &  3.337626E+06  &  1.00006E+00  &  1.61986E+00 \\
2.0 & 1.207510E+02  &  4.656771E+06  &  1.00105E+00  &  1.66647E+00 \\
2.5 & 2.364170E+03  &  6.282161E+06  &  1.00579E+00  &  1.71058E+00 \\
3.0 & 1.834150E+04  &  8.181936E+06  &  1.01832E+00  &  1.75880E+00 \\
3.5 & 8.267880E+04  &  1.035050E+07  &  1.04254E+00  &  1.81336E+00 \\
4.0 & 2.632080E+05  &  1.281267E+07  &  1.08181E+00  &  1.87574E+00 \\
5.0 & 1.400220E+06  &  1.882244E+07  &  1.21974E+00  &  2.03196E+00 \\
6.0 & 4.411610E+06  &  2.665781E+07  &  1.46932E+00  &  2.24833E+00 \\
7.0 & 1.004860E+07  &  3.660870E+07  &  1.89087E+00  &  2.55127E+00 \\
8.0 & 1.829130E+07  &  4.862018E+07  &  2.58102E+00  &  2.97163E+00 \\
9.0 & 2.822910E+07  &  6.224363E+07  &  3.69363E+00  &  3.54622E+00 \\
10.0 & 3.839050E+07  &  7.668148E+07  &  5.47534E+00  &  4.32118E+00 \\

\end{tabular}
\end{ruledtabular}
\end{table*}

\subsubsection{\label{subsec_network}Network Calculations}

During the stellar evolution of a massive star, $^{40}$K can be created both through fusion reactions during carbon and neon burning and via neutron capture reactions in the helium-burning core (s-process) \cite{Woosley2002}. The latter mechanism can also occur in the thermally unstable He inter-shells of asymptotic giant branch (AGB) stars~\cite{Schwarz1965}. The s-process in massive stars is considered to be the dominant mechanism. In any case, due to the existence of these various components, an accurate calculation of stellar yields for $^{40}$K is a rather tricky task. Furthermore, considering the destruction of s-process materials due to the mixing of shells in the stellar interior, the situation becomes even more complicated. Nevertheless, by studying each of the above components individually, any inaccuracies on determining the stellar yields can be reduced. 

We can get an insight into the potential effect of the reaction rates measured in this work on abundance calculations by considering a simple case of nucleosynthesis for $^{40}$K. Here, we focus on the study of s-process nucleosynthesis in the context of the classical model~\cite{Kappel89}, and we investigate the impact of the $^{40}$K(n,p)$^{40}$Ar reaction on the abundance of $^{40}$K.

\begin{figure}
\includegraphics[width=0.48\textwidth]{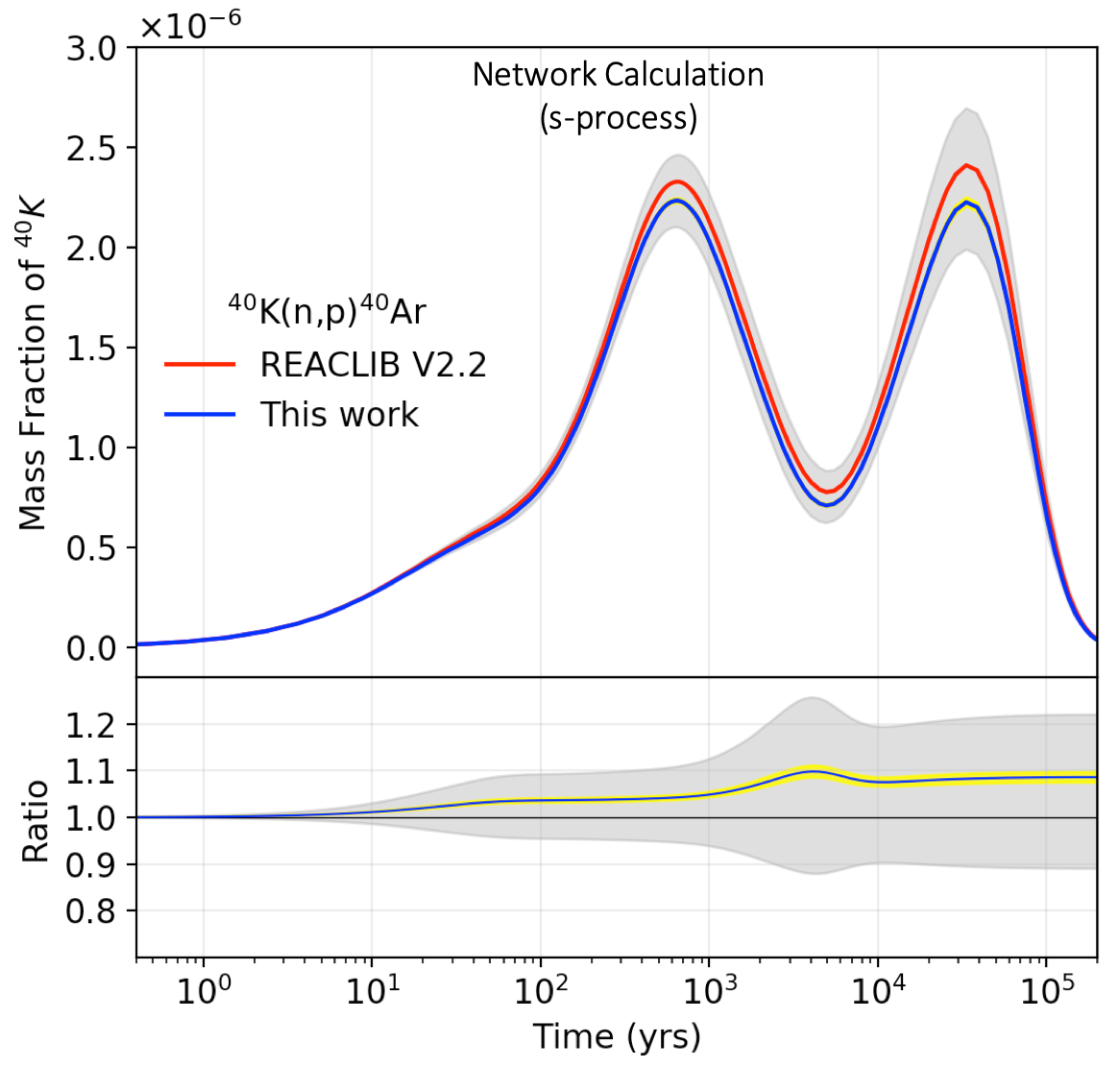}
\caption{Mass fraction of $^{40}$K produced by s-process for different rates of the $^{40}$K(n,p)$^{40}$Ar reaction. The grey error-band highlights the sensitivity of the mass fraction of $^{40}$K on the variation of the library rate by a factor of two. The yellow error-band, represents the upper-limit of the associated uncertainty on the experimental reaction rate, which is $\sim$15\%. The lower panel displays the ratio of the two mass fractions when the numerator of the ratio corresponds to the calculations using the REACLIB rate. The new experimentally constrained rate suggests a factor of 1.39 increase in the destruction rate of $^{40}$K, and a close to 10\% reduction in the production of $^{40}$K after a period of 10$^5$~yrs. }
\label{fig_network}
\end{figure}

Single-zone network calculations were performed using the open-source code NucnetTools~\cite{NucnetTools}. For the initial zone-composition, we considered the solar abundances up to $^{58}$Fe~\cite{Lodders2010}. The temperature and mass density of the zone were set to 0.3~GK and 1000~g/$cm^3$, respectively, while during the calculation, the neutron density was fixed to $10^8$~neutrons/$cm^3$. These conditions were derived based on the classical s-process model approach~\cite{Kappel89,Kappel90}. The reaction network included all the relevant (n,$\gamma$), (n,p), (n,$\alpha$), $\beta^+$, and $\beta^-$ reactions, while the corresponding thermonuclear reactions rates were retrieved from the JINA REACLIB library. In this calculation, the system evolved for $\sim$10$^{5}$~years. 

Fig.~\ref{fig_network}, shows the time evolution of the mass fraction of $^{40}$K according to the above network calculation.
During the evolution of the system, the mass fraction of $^{40}$K peaks at two different times. That happened because the reaction flow through $^{40}$K depends on the abundance evolution of the lighter elements in the reaction chain.  The latter is affected by the initial zone composition, which adds an extra level of complexity in the calculation. To investigate the impact of the $^{40}$K(n,p)$^{40}$Ar reaction on the final $^{40}$K yields, we repeated the calculation using the extracted thermonuclear rate from this study. The corresponding rate at 0.3~GK, was obtained after extrapolating the calculated data for temperatures below 0.4~GK, as described in the previous section. Based on the trend of the extrapolated curve, the induced error due to the extrapolation at 0.3~GK is well within the experimental error of 15\% (see Fig.~\ref{fig_astro_rate_np}). The yellow error-band in Fig.~\ref{fig_network} represents the upper-limits of this error. 

The results of this network calculation show that the abundance of $^{40}$K produced by s-process within a period of time of 10$^5$~years is overestimated by up-to 8.5$\pm$1.2\% when the REACLIB library rate is used. Previous studies on stellar nucleosynthesis~\cite{Hoffman1999}, revealed sensitivity of the order of 20--30\% for the stellar yields of $^{40}$K depending on the choice of theoretical reaction rate. The result was based on a comparison between the rate currently adopted by JINA REACLIB and the previously recommended calculation \cite{Thielemann1996}. Those yield variations were linked to the individual reaction rates that are responsible for the formation and destruction of $^{40}$K. It is generally considered in the community that Hauser-Feshbach calculations are supposed to be accurate within approximately a factor of two. In the present study, we provide an experimental rate for the $^{40}$K(n,p)$^{40}$Ar reaction. The well-defined uncertainty of the order of 15\% for the experimental rate at the energies of interest is a significant improvement over the typical factor-of-two uncertainty of the calculated rate. The reduction in uncertainty is illustrated in Fig.~\ref{fig_network}. Even though many factors must be taken into account to obtain accurate final stellar yields, our measurement provides a significant constraint the uncertainties associated with the destruction rate of $^{40}$K and can be used to inform future studies of the relevant s-process component of $^{40}$K nucleosynthesis.

\section{Conclusion} \label{sec_conclusion}

We measured the total and partial cross-sections of the reaction $^{40}$Ar(p,n)$^{40}$K  for incident proton beam energies between 3.0 - 4.0~MeV. We found that all the cross-sections vary considerably as a function of energy within our experimental resolution. In individual neutron channels, the shapes of the angular distributions were affected by statistical fluctuations. These variations in the cross-section were found to be consistent with theoretical predictions and previous work available in the literature.

To reproduce the experimental data using the Hauser-Feshbach theory, we extracted an improved set of neutron, proton, and alpha transmission coefficients from statistical model calculations. Based on these optimum transmission coefficients, improved astrophysical reaction rates were calculated for the $^{40}$Ar(p,n)$^{40}$K and $^{40}$K(n,p)$^{40}$Ar reactions. 

The new reaction rates were found to be considerably higher than the rates currently available in REACLIB. Using the experimentally constrained rates, we performed network calculations to investigate the implications of this result on s-process nucleosynthesis of $^{40}$K. We found that the abundance of $^{40}$K can be overestimated by $\sim$10\% when using the REACLIB rate in network calculations. The experimental measurement removes a significant portion of the previous theoretical uncertainty on the $^{40}$K yields from stellar evolution calculations. 

The initial mass fraction of $^{40}$K during the formation of a cosmochemically earth-like exoplanet affects the time evolution of the radiogenic heating of the planet. The heating rate has implications on the planet's geological activity, greenhouse gas recycling in the planet's atmosphere, and eventually to the planet's habitability. Currently, age-dependent predictions for such crucial phenomena are made by using galactic chemical evolution (GCE) models. The results from our study can improve the predictions of stellar yields of $^{40}$K at the time of the planet's formation, a critical input parameter for such GCE models. 

Future measurements with proton energies below 2.5~MeV would provide useful information for the cross-sections inside the Gamow-Window for stellar temperatures. Such studies would further constrain the rate of the $^{40}$K(n,p)$^{40}$Ar reaction during stellar nucleosynthesis. Also, measurements on the $^{40}$K(n,$\alpha$)$^{37}$Cl reaction at astrophysical energies would be particularly useful. Constraining all the relevant reaction channels which are responsible for the destruction of $^{40}$K during stellar evolution, will further improve the accuracy of the astrophysical models in predicting the stellar yields of $^{40}$K.

\section*{Acknowledgments} \label{sec_ackow}

We would like to thank Don Carter for his invaluable help in setting up the data acquisition system for this experiment and Devon Jacobs for his help with the gas-cell target and the technical support. 
This material is based upon work supported by the U.S. Department of Energy, Office of Science, under Award Number DE-SC0014285. The authors from Central Michigan University would like to acknowledge support by the College of Science and Engineering of Central Michigan University. The authors from Ohio University would like to acknowledge support from the DOE office of science under grants DE-FG02-88ER40387 and DE-SC0019042, and DOE National Nuclear Security Administration (NNSA) grants DE-NA0003909 and DE-NA0003883.

\nocite{*}

\bibliography{references}% Produces the bibliography via BibTeX.

\newpage

\begin{table*}[h]
\caption{ Local neutron transmission coefficients based on the experimental cross sections of this study. All energies are given in the lab system.}
\label{results_transmission_coeff_neut}
\begin{ruledtabular}
\begin{tabular}{ccccccc}

 \multicolumn{7}{c}{  }\\ 
 \multicolumn{7}{c}{ Neutron transmission coefficients for $L - 1/2$}\\ 
   E$_n$ (MeV) & L=0  & L=1 & L=2 & L=3 & L=4 & L=5  \\ \hline
  0.00103  & 0.000000E+00 & 1.578316E-05 & 1.653768E-09 & 6.402627E-13 & 1.363284E-18 & 1.841157E-23  \\
 0.00205  & 0.000000E+00 & 4.413720E-05 & 9.376920E-09 & 7.318125E-12 & 3.087324E-17 & 8.333415E-22  \\
 0.00513  & 0.000000E+00 & 1.705860E-04 & 9.308070E-08 & 1.841994E-10 & 1.911204E-15 & 1.288359E-19  \\
 0.01025  & 0.000000E+00 & 4.695760E-04 & 5.282739E-07 & 2.119203E-09 & 4.326633E-14 & 5.824719E-18  \\
 0.0205  & 0.000000E+00 & 1.274088E-03 & 3.001338E-06 & 2.449467E-08 & 9.805320E-13 & 2.635632E-16  \\
 0.05126  & 0.000000E+00 & 4.564960E-03 & 2.987100E-05 & 6.280974E-07 & 6.076035E-11 & 4.070241E-14  \\
 0.10252  & 0.000000E+00 & 1.126616E-02 & 1.699533E-04 & 7.368237E-06 & 1.377837E-09 & 1.838385E-12  \\
 0.20505  & 0.000000E+00 & 2.537180E-02 & 9.681660E-04 & 8.716788E-05 & 3.126681E-08 & 8.286336E-11  \\
 0.30757  & 0.000000E+00 & 3.848940E-02 & 2.681235E-03 & 3.697029E-04 & 1.942551E-07 & 7.672761E-10  \\
 0.4101  & 0.000000E+00 & 5.019040E-02 & 5.527494E-03 & 1.024452E-03 & 7.103925E-07 & 3.718008E-09  \\
 0.51262  & 0.000000E+00 & 6.056400E-02 & 9.689940E-03 & 2.237274E-03 & 1.942902E-06 & 1.263204E-08  \\
 0.61514  & 0.000000E+00 & 6.980360E-02 & 1.532862E-02 & 4.186719E-03 & 4.422807E-06 & 3.429207E-08  \\
 0.71767  & 0.000000E+00 & 7.809360E-02 & 2.258091E-02 & 7.021458E-03 & 8.871174E-06 & 7.974504E-08  \\
 0.82019  & 0.000000E+00 & 8.558600E-02 & 3.156129E-02 & 1.084095E-02 & 1.621881E-05 & 1.655658E-07  \\
 0.92272  & 0.000000E+00 & 9.242400E-02 & 4.236039E-02 & 1.568691E-02 & 2.763027E-05 & 3.152817E-07  \\
 1.02524  & 0.000000E+00 & 9.870480E-02 & 5.504076E-02 & 2.154042E-02 & 4.451850E-05 & 5.607504E-07  \\
 1.12776  & 0.000000E+00 & 1.045280E-01 & 6.963156E-02 & 2.833398E-02 & 6.857118E-05 & 9.438030E-07  \\
 1.23029  & 0.000000E+00 & 1.099552E-01 & 8.613774E-02 & 3.596112E-02 & 1.017729E-04 & 1.517850E-06  \\
 1.33281  & 0.000000E+00 & 1.150468E-01 & 1.045215E-01 & 4.429035E-02 & 1.464129E-04 & 2.349396E-06  \\
 1.43534  & 0.000000E+00 & 1.198464E-01 & 1.247211E-01 & 5.318316E-02 & 2.051343E-04 & 3.520152E-06  \\
 1.53786  & 0.000000E+00 & 1.244076E-01 & 1.466091E-01 & 6.251589E-02 & 2.809071E-04 & 5.128164E-06  \\
 1.64038  & 0.000000E+00 & 1.287436E-01 & 1.700703E-01 & 7.215219E-02 & 3.771117E-04 & 7.290351E-06  \\
 1.74291  & 0.000000E+00 & 1.329004E-01 & 1.949211E-01 & 8.200053E-02 & 4.975398E-04 & 1.014462E-05  \\

 \multicolumn{7}{c}{  }\\ 
 \multicolumn{7}{c}{ Neutron transmission coefficients for $L + 1/2$}\\ 
   E$_n$ (MeV) & L=0  & L=1 & L=2 & L=3 & L=4 & L=5  \\ \hline
  0.00103  & 1.242388E-02 & 6.835040E-06 & 2.847816E-09 & 7.547760E-14 & 3.545811E-18 & 2.154303E-23  \\
 0.00205  & 1.759268E-02 & 1.919740E-05 & 1.618092E-08 & 8.512029E-13 & 8.058267E-17 & 9.753210E-22  \\
 0.00513  & 2.782444E-02 & 7.482640E-05 & 1.612314E-07 & 2.092275E-11 & 5.020137E-15 & 1.508562E-19  \\
 0.01025  & 3.923760E-02 & 2.079608E-04 & 9.186570E-07 & 2.351907E-10 & 1.143711E-13 & 6.823512E-18  \\
 0.0205  & 5.505840E-02 & 5.723440E-04 & 5.246820E-06 & 2.640924E-09 & 2.613213E-12 & 3.089556E-16  \\
 0.05126  & 8.499480E-02 & 2.117600E-03 & 5.277726E-05 & 6.435288E-08 & 1.644264E-10 & 4.777362E-14  \\
 0.10252  & 1.159928E-01 & 5.451800E-03 & 3.043665E-04 & 7.157097E-07 & 3.794940E-09 & 2.161233E-12  \\
 0.20505  & 1.545488E-01 & 1.312600E-02 & 1.775286E-03 & 7.879284E-06 & 8.855676E-08 & 9.768060E-11  \\
 0.30757  & 1.799864E-01 & 2.099636E-02 & 5.027949E-03 & 3.179358E-05 & 5.639130E-07 & 9.067500E-10  \\
 0.4101  & 1.988204E-01 & 2.861168E-02 & 1.059579E-02 & 8.510112E-05 & 2.110914E-06 & 4.404627E-09  \\
 0.51262  & 2.135768E-01 & 3.584680E-02 & 1.897911E-02 & 1.819089E-04 & 5.905350E-06 & 1.500111E-08  \\
 0.61514  & 2.255716E-01 & 4.268880E-02 & 3.065481E-02 & 3.373515E-04 & 1.374435E-05 & 4.082211E-08  \\
 0.71767  & 2.355648E-01 & 4.916000E-02 & 4.605975E-02 & 5.672601E-04 & 2.817765E-05 & 9.516060E-08  \\
 0.82019  & 2.440648E-01 & 5.529080E-02 & 6.556860E-02 & 8.878185E-04 & 5.264604E-05 & 1.980558E-07  \\
 0.92272  & 2.513944E-01 & 6.112360E-02 & 8.945955E-02 & 1.315647E-03 & 9.163440E-05 & 3.780819E-07  \\
 1.02524  & 2.578060E-01 & 6.668760E-02 & 1.178856E-01 & 1.867239E-03 & 1.508256E-04 & 6.741207E-07  \\
 1.12776  & 2.634584E-01 & 7.202080E-02 & 1.508193E-01 & 2.559384E-03 & 2.372562E-04 & 1.137474E-06  \\
 1.23029  & 2.685048E-01 & 7.714560E-02 & 1.880757E-01 & 3.408570E-03 & 3.595527E-04 & 1.833984E-06  \\
 1.33281  & 2.730460E-01 & 8.208840E-02 & 2.292381E-01 & 4.431051E-03 & 5.280201E-04 & 2.846052E-06  \\
 1.43534  & 2.771696E-01 & 8.686760E-02 & 2.737251E-01 & 5.642892E-03 & 7.549794E-04 & 4.275450E-06  \\
 1.53786  & 2.809160E-01 & 9.150640E-02 & 3.206907E-01 & 7.059771E-03 & 1.054575E-03 & 6.244965E-06  \\
 1.64038  & 2.843664E-01 & 9.601280E-02 & 3.693015E-01 & 8.696475E-03 & 1.443627E-03 & 8.901882E-06  \\
 1.74291  & 2.875468E-01 & 1.004096E-01 & 4.185261E-01 & 1.056843E-02 & 1.941003E-03 & 1.242081E-05  \\

\end{tabular}
\end{ruledtabular}
\end{table*}

\begingroup
\squeezetable
\begin{table*}[h]
\caption{ Local proton transmission coefficients based on the experimental cross sections of this study. All energies are given in the lab system.}
\label{results_transmission_coeff_prot}
\begin{ruledtabular}
\begin{tabular}{cccccccc}

 \multicolumn{8}{c}{  }\\ 
 \multicolumn{8}{c}{ Proton transmission coefficients for $L - 1/2$}\\ 
   E$_p$ (MeV) & L=0  & L=1 & L=2 & L=3 & L=4 & L=5 & L=6  \\ \hline
 0.10252  & 0.000000E+00 & 7.722550E-21 & 1.056870E-22 & 3.643920E-24 & 1.867750E-26 & 9.556740E-29 & 3.493620E-31  \\
 0.20504  & 0.000000E+00 & 1.040800E-13 & 1.387720E-15 & 5.483260E-17 & 3.219900E-19 & 2.039370E-21 & 9.788000E-24  \\
 0.30756  & 0.000000E+00 & 1.543280E-10 & 2.021160E-12 & 9.024680E-14 & 5.957430E-16 & 4.496040E-18 & 2.672080E-20  \\
 0.41008  & 0.000000E+00 & 1.223610E-08 & 1.577960E-10 & 7.886010E-12 & 5.763340E-14 & 5.046430E-16 & 3.576190E-18  \\
 0.5126  & 0.000000E+00 & 2.457820E-07 & 3.123750E-09 & 1.734200E-10 & 1.386020E-12 & 1.380600E-14 & 1.136180E-16  \\
 0.61512  & 0.000000E+00 & 2.279510E-06 & 2.856170E-08 & 1.750550E-09 & 1.514740E-11 & 1.690630E-13 & 1.584520E-15  \\
 0.71764  & 0.000000E+00 & 1.300650E-05 & 1.606940E-07 & 1.081650E-08 & 1.004860E-10 & 1.241690E-12 & 1.305600E-14  \\
 0.82016  & 0.000000E+00 & 5.340960E-05 & 6.507640E-07 & 4.789260E-08 & 4.743040E-10 & 6.425770E-12 & 7.490220E-14  \\
 0.92269  & 0.000000E+00 & 1.733160E-04 & 2.084040E-06 & 1.670260E-07 & 1.752780E-09 & 2.582520E-11 & 3.305130E-13  \\
 1.02521  & 0.000000E+00 & 4.717450E-04 & 5.603430E-06 & 4.873800E-07 & 5.391000E-09 & 8.579690E-11 & 1.195920E-12  \\
 1.12773  & 0.000000E+00 & 1.120750E-03 & 1.317470E-05 & 1.239670E-06 & 1.438810E-08 & 2.459030E-10 & 3.708060E-12  \\
 1.23025  & 0.000000E+00 & 2.389710E-03 & 2.786800E-05 & 2.828920E-06 & 3.431290E-08 & 6.265920E-10 & 1.016280E-11  \\
 1.33277  & 0.000000E+00 & 4.663530E-03 & 5.414100E-05 & 5.914060E-06 & 7.469820E-08 & 1.451100E-09 & 2.518880E-11  \\
 1.43529  & 0.000000E+00 & 8.449270E-03 & 9.809940E-05 & 1.150490E-05 & 1.508330E-07 & 3.104990E-09 & 5.743260E-11  \\
 1.53781  & 0.000000E+00 & 1.435620E-02 & 1.677430E-04 & 2.107400E-05 & 2.859910E-07 & 6.217260E-09 & 1.220740E-10  \\
 1.64033  & 0.000000E+00 & 2.305850E-02 & 2.731600E-04 & 3.669260E-05 & 5.140890E-07 & 1.176600E-08 & 2.444020E-10  \\
 1.74285  & 0.000000E+00 & 3.520220E-02 & 4.266930E-04 & 6.116570E-05 & 8.827710E-07 & 2.121220E-08 & 4.647200E-10  \\
 1.84537  & 0.000000E+00 & 5.130770E-02 & 6.430940E-04 & 9.820700E-05 & 1.457040E-06 & 3.666620E-08 & 8.449240E-10  \\
 1.94789  & 0.000000E+00 & 7.164990E-02 & 9.395490E-04 & 1.525980E-04 & 2.323090E-06 & 6.108400E-08 & 1.476910E-09  \\
 2.05041  & 0.000000E+00 & 9.616160E-02 & 1.335840E-03 & 2.303850E-04 & 3.592950E-06 & 9.850910E-08 & 2.493470E-09  \\
 2.25545  & 0.000000E+00 & 1.556010E-01 & 2.519610E-03 & 4.878190E-04 & 7.949920E-06 & 2.356630E-07 & 6.498800E-09  \\
 2.46049  & 0.000000E+00 & 2.227670E-01 & 4.403130E-03 & 9.515820E-04 & 1.613210E-05 & 5.136800E-07 & 1.532670E-08  \\
 2.66553  & 0.000000E+00 & 2.895120E-01 & 7.233160E-03 & 1.735550E-03 & 3.050370E-05 & 1.037500E-06 & 3.329610E-08  \\
 2.87058  & 0.000000E+00 & 3.499550E-01 & 1.129110E-02 & 2.991980E-03 & 5.439960E-05 & 1.966680E-06 & 6.754300E-08  \\
 3.07562  & 0.000000E+00 & 4.015550E-01 & 1.688470E-02 & 4.915030E-03 & 9.234350E-05 & 3.533170E-06 & 1.292760E-07  \\
 3.28066  & 0.000000E+00 & 4.443340E-01 & 2.433930E-02 & 7.740100E-03 & 1.502910E-04 & 6.062470E-06 & 2.353990E-07  \\
 3.4857  & 0.000000E+00 & 4.793600E-01 & 3.398630E-02 & 1.174160E-02 & 2.358790E-04 & 9.996930E-06 & 4.104950E-07  \\
 3.69074  & 0.000000E+00 & 5.081770E-01 & 4.614450E-02 & 1.721700E-02 & 3.586700E-04 & 1.592160E-05 & 6.892260E-07  \\
 3.89578  & 0.000000E+00 & 5.320870E-01 & 6.110850E-02 & 2.447350E-02 & 5.304330E-04 & 2.459290E-05 & 1.119180E-06  \\
 4.10082  & 0.000000E+00 & 5.522950E-01 & 7.911850E-02 & 3.379400E-02 & 7.653390E-04 & 3.696680E-05 & 1.764040E-06  \\

 \multicolumn{8}{c}{  }\\ 
 \multicolumn{8}{c}{ Proton transmission coefficients for $L + 1/2$}\\ 
   E$_p$ (MeV) & L=0  & L=1 & L=2 & L=3 & L=4 & L=5 & L=6  \\ \hline

 0.10252  & 5.828490E-21 & 2.431560E-20 & 1.034850E-22 & 1.197200E-23 & 2.728010E-26 & 1.051510E-28 & 0.000000E+00  \\
 0.20504  & 7.190970E-14 & 3.195740E-13 & 1.385930E-15 & 1.627620E-16 & 4.743590E-19 & 2.246750E-21 & 1.015320E-23  \\
 0.30756  & 9.833320E-11 & 4.549770E-10 & 2.050840E-12 & 2.454330E-13 & 8.844590E-16 & 4.959080E-18 & 2.772480E-20  \\
 0.41008  & 7.212050E-09 & 3.416070E-08 & 1.623850E-10 & 1.977890E-11 & 8.619950E-14 & 5.572540E-16 & 3.711500E-18  \\
 0.5126  & 1.342420E-07 & 6.414340E-07 & 3.256630E-09 & 4.028490E-10 & 2.088080E-12 & 1.526260E-14 & 1.179470E-16  \\
 0.61512  & 1.155250E-06 & 5.493870E-06 & 3.014320E-08 & 3.777840E-09 & 2.298430E-11 & 1.871120E-13 & 1.645300E-15  \\
 0.71764  & 6.123470E-06 & 2.863230E-05 & 1.715840E-07 & 2.173960E-08 & 1.535690E-10 & 1.375810E-12 & 1.356020E-14  \\
 0.82016  & 2.338550E-05 & 1.063400E-04 & 7.027450E-07 & 8.981570E-08 & 7.300830E-10 & 7.127970E-12 & 7.781500E-14  \\
 0.92269  & 7.068430E-05 & 3.095780E-04 & 2.275200E-06 & 2.928160E-07 & 2.717460E-09 & 2.868040E-11 & 3.434550E-13  \\
 1.02521  & 1.795220E-04 & 7.509510E-04 & 6.182940E-06 & 7.999220E-07 & 8.418730E-09 & 9.539330E-11 & 1.243080E-12  \\
 1.12773  & 3.989710E-04 & 1.582820E-03 & 1.468900E-05 & 1.907930E-06 & 2.263240E-08 & 2.737280E-10 & 3.855300E-12  \\
 1.23025  & 7.981920E-04 & 2.985350E-03 & 3.139010E-05 & 4.087950E-06 & 5.436950E-08 & 6.983240E-10 & 1.056920E-11  \\
 1.33277  & 1.467360E-03 & 5.148560E-03 & 6.160000E-05 & 8.034020E-06 & 1.192350E-07 & 1.619170E-09 & 2.620320E-11  \\
 1.43529  & 2.516820E-03 & 8.251300E-03 & 1.127310E-04 & 1.470830E-05 & 2.425550E-07 & 3.468830E-09 & 5.976230E-11  \\
 1.53781  & 4.074610E-03 & 1.244640E-02 & 1.946620E-04 & 2.538960E-05 & 4.633360E-07 & 6.954350E-09 & 1.270620E-10  \\
 1.64033  & 6.283120E-03 & 1.783780E-02 & 3.201070E-04 & 4.169660E-05 & 8.391620E-07 & 1.317740E-08 & 2.544620E-10  \\
 1.74285  & 9.293160E-03 & 2.448740E-02 & 5.048870E-04 & 6.564060E-05 & 1.451890E-06 & 2.378670E-08 & 4.839900E-10  \\
 1.84537  & 1.325990E-02 & 3.239920E-02 & 7.683060E-04 & 9.963090E-05 & 2.414640E-06 & 4.116910E-08 & 8.802220E-10  \\
 1.94789  & 1.833480E-02 & 4.153270E-02 & 1.133300E-03 & 1.465030E-04 & 3.879370E-06 & 6.867460E-08 & 1.539070E-09  \\
 2.05041  & 2.466140E-02 & 5.180730E-02 & 1.626780E-03 & 2.095370E-04 & 6.046100E-06 & 1.108960E-07 & 2.599230E-09  \\
 2.25545  & 4.156120E-02 & 7.530680E-02 & 3.127430E-03 & 3.993960E-04 & 1.358600E-05 & 2.660100E-07 & 6.778730E-09  \\
 2.46049  & 6.471100E-02 & 1.018240E-01 & 5.570070E-03 & 7.043630E-04 & 2.800020E-05 & 5.814270E-07 & 1.599740E-08  \\
 2.66553  & 9.440550E-02 & 1.302230E-01 & 9.324670E-03 & 1.166530E-03 & 5.377690E-05 & 1.177650E-06 & 3.477670E-08  \\
 2.87058  & 1.304320E-01 & 1.595170E-01 & 1.483090E-02 & 1.834770E-03 & 9.741230E-05 & 2.238860E-06 & 7.059610E-08  \\
 3.07562  & 1.720890E-01 & 1.889060E-01 & 2.259030E-02 & 2.763740E-03 & 1.679510E-04 & 4.034190E-06 & 1.352180E-07  \\
 3.28066  & 2.182720E-01 & 2.178770E-01 & 3.315080E-02 & 4.014110E-03 & 2.775900E-04 & 6.943380E-06 & 2.464030E-07  \\
 3.4857  & 2.676800E-01 & 2.460070E-01 & 4.709060E-02 & 5.649860E-03 & 4.423780E-04 & 1.148570E-05 & 4.300200E-07  \\
 3.69074  & 3.188820E-01 & 2.731220E-01 & 6.497230E-02 & 7.739510E-03 & 6.828310E-04 & 1.835180E-05 & 7.225900E-07  \\
 3.89578  & 3.705700E-01 & 2.990770E-01 & 8.731640E-02 & 1.035250E-02 & 1.024780E-03 & 2.844090E-05 & 1.174330E-06  \\
 4.10082  & 4.215270E-01 & 3.238780E-01 & 1.145220E-01 & 1.356080E-02 & 1.499830E-03 & 4.289660E-05 & 1.852570E-06  \\

\end{tabular}
\end{ruledtabular}
\end{table*}
\endgroup

\begingroup
\squeezetable
\begin{table*}[h]
\caption{ Local alpha transmission coefficients based on the experimental cross sections of this study. All energies are given in the lab system.}
\label{results_transmission_coeff_alp}
\begin{ruledtabular}
\begin{tabular}{ccccccc}

 \multicolumn{7}{c}{  }\\ 
 \multicolumn{7}{c}{ Alpha transmission coefficients }\\ 
   E$_\alpha$ (MeV) & L=0  & L=1 & L=2 & L=3 & L=4 & L=5  \\ \hline
  0.22166  & 0.000000E+00 & 0.000000E+00 & 0.000000E+00 & 0.000000E+00 & 0.000000E+00 & 0.000000E+00  \\
 0.33248  & 3.198360E-38 & 4.461578E-38 & 0.000000E+00 & 0.000000E+00 & 0.000000E+00 & 0.000000E+00  \\
 0.44331  & 1.820401E-31 & 2.582580E-31 & 4.308612E-32 & 0.000000E+00 & 0.000000E+00 & 0.000000E+00  \\
 0.55414  & 7.233974E-27 & 1.038543E-26 & 1.716227E-27 & 8.005910E-28 & 7.811914E-29 & 1.283869E-29  \\
 0.66497  & 1.761492E-23 & 2.545972E-23 & 4.189240E-24 & 2.009341E-24 & 1.929283E-25 & 3.280420E-26  \\
 0.77579  & 7.426188E-21 & 1.074988E-20 & 1.770536E-21 & 8.701792E-22 & 8.246986E-23 & 1.450445E-23  \\
 0.88662  & 9.574400E-19 & 1.380911E-18 & 2.288594E-19 & 1.148160E-19 & 1.077828E-20 & 1.959817E-21  \\
 0.99745  & 5.300482E-17 & 7.578560E-17 & 1.270361E-17 & 6.478978E-18 & 6.047404E-19 & 1.135902E-19  \\
 1.10828  & 1.566376E-15 & 2.209394E-15 & 3.764420E-16 & 1.943370E-16 & 1.810864E-17 & 3.509616E-18  \\
 1.21911  & 2.861100E-14 & 3.962838E-14 & 6.895482E-15 & 3.587452E-15 & 3.351128E-16 & 6.691190E-17  \\
 1.32993  & 3.579466E-13 & 4.847348E-13 & 8.651984E-14 & 4.516226E-14 & 4.246990E-15 & 8.720030E-16  \\
 1.44076  & 3.306270E-12 & 4.360026E-12 & 8.015722E-13 & 4.179670E-13 & 3.973376E-14 & 8.370560E-15  \\
 1.55159  & 2.383018E-11 & 3.048826E-11 & 5.795328E-12 & 3.005882E-12 & 2.900436E-13 & 6.253500E-14  \\
 1.66242  & 1.398276E-10 & 1.729726E-10 & 3.411342E-11 & 1.752854E-11 & 1.723469E-12 & 3.792470E-13  \\
 1.77325  & 6.903116E-10 & 8.231256E-10 & 1.689648E-10 & 8.567790E-11 & 8.615838E-12 & 1.929206E-12  \\
 1.88407  & 2.942720E-09 & 3.372798E-09 & 7.226912E-10 & 3.603380E-10 & 3.718902E-11 & 8.446988E-12  \\
 1.9949  & 1.106387E-08 & 1.215779E-08 & 2.726416E-09 & 1.332219E-09 & 1.415641E-10 & 3.251314E-11  \\
 2.10573  & 3.731200E-08 & 3.921918E-08 & 9.226624E-09 & 4.404576E-09 & 4.833312E-10 & 1.118863E-10  \\
 2.21656  & 1.144603E-07 & 1.148387E-07 & 2.840420E-08 & 1.320946E-08 & 1.500970E-09 & 3.490982E-10  \\
 2.43821  & 8.477238E-07 & 7.705280E-07 & 2.118807E-07 & 9.280370E-08 & 1.138927E-08 & 2.649504E-09  \\
 2.65987  & 4.835534E-06 & 3.955006E-06 & 1.217335E-06 & 4.977566E-07 & 6.652822E-08 & 1.530459E-08  \\
 2.88152  & 2.322782E-05 & 1.673465E-05 & 5.900202E-06 & 2.187612E-06 & 3.286492E-07 & 7.220818E-08  \\
 3.10318  & 1.106585E-04 & 6.655748E-05 & 2.879404E-05 & 8.903004E-06 & 1.674666E-06 & 3.187184E-07  \\
 3.32484  & 4.191572E-04 & 2.251502E-04 & 1.123562E-04 & 3.123230E-05 & 6.884438E-06 & 1.217359E-06  \\
 3.54649  & 1.322878E-03 & 6.692532E-04 & 3.670018E-04 & 9.681320E-05 & 2.387220E-05 & 4.090966E-06  \\
 3.76815  & 3.598562E-03 & 1.784185E-03 & 1.036906E-03 & 2.697838E-04 & 7.201810E-05 & 1.229384E-05  \\
 3.9898  & 8.657484E-03 & 4.327994E-03 & 2.597848E-03 & 6.849458E-04 & 1.934425E-04 & 3.351150E-05  \\
 4.21146  & 1.879280E-02 & 9.659210E-03 & 5.885968E-03 & 1.602033E-03 & 4.711410E-04 & 8.388798E-05  \\
 4.43311  & 3.736172E-02 & 2.000064E-02 & 1.224485E-02 & 3.483458E-03 & 1.055641E-03 & 1.948320E-04  \\
 4.98725  & 1.511708E-01 & 9.230650E-02 & 5.645376E-02 & 1.849551E-02 & 5.914502E-03 & 1.217843E-03  \\
 5.54139  & 4.138728E-01 & 2.922326E-01 & 1.828867E-01 & 7.079644E-02 & 2.392412E-02 & 5.556452E-03  \\
 6.09553  & 8.093492E-01 & 6.513056E-01 & 4.375360E-01 & 2.043406E-01 & 7.513814E-02 & 1.979912E-02  \\

 \multicolumn{7}{c}{  }\\ 
 %\multicolumn{7}{c}{ Alpha transmission coefficients ...continue}\\ 
   E$_\alpha$ (MeV) & L=6  & L=7 & L=8 & L=9 & L=10 & L=11  \\ \hline
 
  0.22166  & 0.000000E+00 & 0.000000E+00 & 0.000000E+00 & 0.000000E+00 & 0.000000E+00 & 0.000000E+00  \\
 0.33248  & 0.000000E+00 & 0.000000E+00 & 0.000000E+00 & 0.000000E+00 & 0.000000E+00 & 0.000000E+00  \\
 0.44331  & 0.000000E+00 & 0.000000E+00 & 0.000000E+00 & 0.000000E+00 & 0.000000E+00 & 0.000000E+00  \\
 0.55414  & 1.095406E-30 & 9.621986E-32 & 6.748104E-33 & 4.118114E-34 & 0.000000E+00 & 0.000000E+00  \\
 0.66497  & 2.804208E-27 & 2.551208E-28 & 1.859037E-29 & 1.194794E-30 & 6.655022E-32 & 3.253800E-33  \\
 0.77579  & 1.240380E-24 & 1.167386E-25 & 8.814960E-27 & 5.948382E-28 & 3.507768E-29 & 1.830981E-30  \\
 0.88662  & 1.674726E-22 & 1.628752E-23 & 1.271246E-24 & 8.982336E-26 & 5.588132E-27 & 3.100724E-28  \\
 0.99745  & 9.692056E-21 & 9.730798E-22 & 7.831802E-23 & 5.779972E-24 & 3.781404E-25 & 2.221846E-26  \\
 1.10828  & 2.989030E-19 & 3.095224E-20 & 2.563110E-21 & 1.971286E-22 & 1.352252E-23 & 8.385168E-25  \\
 1.21911  & 5.688672E-18 & 6.070548E-19 & 5.161200E-20 & 4.128058E-21 & 2.961354E-22 & 1.931965E-23  \\
 1.32993  & 7.404166E-17 & 8.135622E-18 & 7.087740E-19 & 5.884076E-20 & 4.403586E-21 & 3.014000E-22  \\
 1.44076  & 7.104856E-16 & 8.031848E-17 & 7.157106E-18 & 6.156348E-19 & 4.795780E-20 & 3.434948E-21  \\
 1.55159  & 5.312626E-15 & 6.173794E-16 & 5.617744E-17 & 4.998598E-18 & 4.044810E-19 & 3.024626E-20  \\
 1.66242  & 3.229776E-14 & 3.854906E-15 & 3.576562E-16 & 3.286954E-17 & 2.757546E-18 & 2.148254E-19  \\
 1.77325  & 1.649991E-13 & 2.020724E-14 & 1.909151E-15 & 1.809619E-16 & 1.571185E-17 & 1.272680E-18  \\
 1.88407  & 7.269790E-13 & 9.125930E-14 & 8.770234E-15 & 8.562466E-16 & 7.681168E-17 & 6.457418E-18  \\
 1.9949  & 2.821742E-12 & 3.626612E-13 & 3.541978E-14 & 3.557378E-15 & 3.292124E-16 & 2.867568E-17  \\
 2.10573  & 9.813254E-12 & 1.289671E-12 & 1.279172E-13 & 1.320046E-14 & 1.258437E-15 & 1.133931E-16  \\
 2.21656  & 3.101054E-11 & 4.161564E-12 & 4.189746E-13 & 4.437422E-14 & 4.352040E-15 & 4.050662E-16  \\
 2.43821  & 2.429570E-10 & 3.384238E-11 & 3.507438E-12 & 3.900160E-13 & 4.033788E-14 & 3.989700E-15  \\
 2.65987  & 1.460089E-09 & 2.097163E-10 & 2.236982E-11 & 2.600840E-12 & 2.824800E-13 & 2.954556E-14  \\
 2.88152  & 7.414550E-09 & 1.078246E-09 & 1.198980E-10 & 1.450198E-11 & 1.650128E-12 & 1.817246E-13  \\
 3.10318  & 3.953334E-08 & 5.561226E-09 & 6.762910E-10 & 8.446900E-11 & 1.008737E-11 & 1.165127E-12  \\
 3.32484  & 1.711567E-07 & 2.393116E-08 & 3.094740E-09 & 3.991592E-10 & 4.975784E-11 & 6.007584E-12  \\
 3.54649  & 6.288942E-07 & 8.857596E-08 & 1.201682E-08 & 1.598997E-09 & 2.072880E-10 & 2.608232E-11  \\
 3.76815  & 2.020953E-06 & 2.890316E-07 & 4.081660E-08 & 5.596602E-09 & 7.522812E-10 & 9.838026E-11  \\
 3.9898  & 5.805228E-06 & 8.476490E-07 & 1.239643E-07 & 1.749508E-08 & 2.432320E-09 & 3.297910E-10  \\
 4.21146  & 1.516174E-05 & 2.268904E-06 & 3.423750E-07 & 4.968216E-08 & 7.128682E-09 & 9.999176E-10  \\
 4.43311  & 3.648744E-05 & 5.611298E-06 & 8.713056E-07 & 1.298744E-07 & 1.919575E-08 & 2.779898E-09  \\
 4.98725  & 2.450360E-04 & 4.061618E-05 & 6.719504E-06 & 1.067046E-06 & 1.686384E-07 & 2.625260E-08  \\
 5.54139  & 1.190103E-03 & 2.133419E-04 & 3.734698E-05 & 6.287974E-06 & 1.053903E-06 & 1.747825E-07  \\
 6.09553  & 4.524960E-03 & 8.769398E-04 & 1.619449E-04 & 2.879096E-05 & 5.085234E-06 & 8.919482E-07  \\

\end{tabular}
\end{ruledtabular}
\end{table*}
\endgroup

\end{document}